\documentclass[fleqn,usenatbib]{mnras}

\usepackage{newtxtext,newtxmath}

\usepackage[T1]{fontenc}

\DeclareRobustCommand{\VAN}[3]{#2}
\let\VANthebibliography\thebibliography
\def\thebibliography{\DeclareRobustCommand{\VAN}[3]{##3}\VANthebibliography}

\usepackage{graphicx}	
\usepackage{amsmath}	

\newcommand{\oiii}{[O~{\sc iii}]}
\newcommand{\lya}{Ly$\alpha$}
\newcommand{\ha}{H$\alpha$}
\newcommand{\hb}{H$\beta$}

\newcommand{\xion}{$\xi_{\rm{ion}}$}
\newcommand{\xionnofesc}{$\xi_{\rm{ion},0}$}

\newcommand{\texp}{T$_{\mathrm{exp}}$}

\title[Scatter of the SFMS]{Bursting at the seams: the star-forming main sequence and its scatter at z=3-9 using NIRCam photometry from JADES}

\author[Simmonds \& Tacchella et al.]{C. Simmonds,$^{1,2}$\thanks{E-mail: cs2210@cam.ac.uk}
S. Tacchella,$^{1,2}$\thanks{E-mail: st578@cam.ac.uk}
W. McClymont,$^{1,2}$
E. Curtis-Lake,$^{3}$
F. D'Eugenio,$^{1,2}$
K. Hainline,$^{4}$
\newauthor
B.~D. Johnson,$^{5}$
A. Kravtsov,$^{6,7,8}$
D. Puskás,$^{1,2}$
B. Robertson,$^{9}$
A. Stoffers,$^{1,2}$
C. Willott,$^{10}$
W.~M. Baker,$^{11}$
\newauthor
V.~A. Belokurov,$^{12}$
R. Bhatawdekar,$^{13}$
A.~J. Bunker,$^{14}$
S. Carniani,$^{15}$
J. Chevallard,$^{14}$
M. Curti,$^{16}$
\newauthor
Q. Duan,$^{1,2}$
J.~M. Helton,$^{4}$
Z. Ji,$^{4}$
T.~J. Looser,$^{5}$
R. Maiolino,$^{1,2,17}$
M.~V. Maseda,$^{18}$
I. Shivaei,$^{19}$
\newauthor
and C.~C. Williams$^{20}$
\\
\\
\emph{\normalsize Affiliations are listed at the end of the paper.}
}

\date{Accepted XXX. Received YYY; in original form ZZZ}
\pubyear{\the\year{}}

\begin{document}
\label{firstpage}
\pagerange{\pageref{firstpage}--\pageref{lastpage}}
\maketitle

\begin{abstract}
We present a comprehensive study of the star-forming main sequence (SFMS) and its scatter at redshifts $3 \leq z \leq 9$, using NIRCam photometry from the JADES survey in the GOODS-S and GOODS-N fields. Our analysis is based on a sample of galaxies that is stellar mass complete down to $\log \left(M_{\star}/M_{\odot}\right) \approx 8.1$. The redshift evolution of the SFMS at an averaging timescale of 10 Myr follows a relation, quantified by the specific star-formation rates (sSFR$_{10}$), of $\mathrm{sSFR}\propto(1+z)^{\mu}$ with $\mu = 2.30^{+0.03}_{-0.01}$, in good agreement with theoretical predictions and the specific mass accretion rate of dark matter halos. We find that the SFMS normalisation varies in a complex way with the SFR averaging timescale, reflecting the combined effects of bursty star formation and rising star formation histories (SFHs). We quantify the scatter of the SFMS, revealing that it decreases with longer SFR averaging timescales, from $\sigma_{\rm{int}} \approx 0.4-0.5~\mathrm{dex}$ at 10 Myr to $\sigma_{\rm{int}} \approx 0.2~\mathrm{dex}$ at 100 Myr, indicating that shorter-term fluctuations dominate the scatter, although long-term variations in star formation activity are also present. Our findings suggest that bursty SFHs are more pronounced at lower stellar masses. Furthermore, we explore the implications of our results for the observed over-abundance of UV-bright galaxies at $z > 10$, concluding that additional mechanisms, such as top-heavy initial mass functions, increased star-formation efficiencies, or increased burstiness in star formation are needed to explain these observations. Finally, we emphasize the importance of accurate stellar mass completeness limits when fitting the SFMS, especially for galaxies with bursty SFHs.
\end{abstract}




\section{Introduction}

The cosmic star formation history (SFH) traces the evolution of the star formation rate (SFR) density across cosmic time, capturing the collective star formation activity of all galaxies from a few million years after the Big Bang to the present, with a peak at redshift $z \approx 2$ \citep[see][and references within]{Madau2014}. While this global history provides critical insight into the overall star formation process in the Universe, the SFHs of individual galaxies reveal the underlying physical mechanisms driving galaxy formation and evolution. These individual histories vary significantly, reflecting the complex interplay between gas accretion, star formation, and feedback processes \citep{Dekel2013,Tacchella2016, Iyer2020,McClymont2025}. Understanding how these diverse SFHs contribute to the cosmic SFR density is key to unravelling the detailed processes of galaxy assembly over time.

An important scaling relation to study SFHs of galaxies is the star-forming main sequence (SFMS), which is the tight relation between SFR and stellar mass that has been widely established for star-forming galaxies up to $z\sim3$ \citep[including, but not limited to][]{Brinchmann2004,Reddy2006,Daddi2007,Noeske2007,Salim2007,Whitaker2012,Speagle2014,Popesso2023}. The SFMS is usually described as a linear relation in logarithmic space between SFR (or specific SFR; sSFR$\equiv$SFR/M$_{\star}$) and stellar mass (M$_{\star}$). On average, studies show that at fixed M$_{\star}$, SFR (and sSFR) increase as a function of redshift, although we note that the specific shape of this relation highly depends on the sample selection and the techniques used to estimate SFRs \citep{Katsianis2016,Davies2019}. With a relatively narrow SFMS scatter of $\sigma \approx 0.3$ dex, the increase in SFMS normalisation from $z \approx 0$ to $z \approx 3$ can be primarily attributed to higher gas fractions in galaxies, with star-formation efficiency and galaxy-galaxy mergers playing sub-dominant roles in driving higher SFRs \citep{Bouche2010,Daddi2010,Rodighiero2011,Tacconi2020}.

Although several studies have demonstrated the presence of the SFMS extending to the Epoch of Reionisation (EoR; $z \approx 5-9$; \citealt{Daddi2010,Speagle2014,Salmon2015,Schreiber2015,Tasca2015,Santini2017,Popesso2023}), until the advent of the James Webb Space Telescope (JWST) its exact shape and scatter remained uncertain due to the observational limitation of mainly probing the rest-frame ultra-violet (UV) emission at these redshifts. Theoretically, the SFMS at $z = 3-9$ is predicted by simple models of galaxy formation, where the SFRs of galaxies are closely tied to dark matter accretion \citep{Rodriguez-Puebla2016,Tacchella2018}. The key insight, however, comes from examining the distribution of galaxies around the SFMS, characterised by its scatter, as this encodes the underlying physical processes driving star formation on global \citep{Gladders2013,Abramson2014,Kelson2014,Tacchella2016, Caplar2019,Matthee2019,Enci2020,Iyer2020,McClymont2025,Wan2025} and spatially resolved scales, giving rise to bulges and disks \citep{Zolotov2015,Tacchella2016b,Appleby2020,Emami2021,Hopkins2023,Lapiner2023,Cenci2024,McClymont2025quench}. 

Burstiness -- the variability of the SFR on short timescales (relative to the dynamical timescale of galaxies, e.g. a few tens of Myr) -- is predicted to increase toward lower-mass galaxies and higher redshifts. In low-mass (M$_{\star}<10^9~\text{M}_{\odot}$) galaxies, bursty star formation potentially arises from episodic stellar feedback processes that expel gas from the interstellar medium \citep[ISM; ][]{Hopkins2014,Hayward2017,Hopkins2023,Dome2024}, as well as from environmental effects such as galaxy-galaxy interactions \citep{DiMatteo2007,Teyssier2010,Asada2024,Duan2024}. At higher redshifts, burstiness is expected across all galaxy masses and is likely due to highly variable gas accretion and short equilibrium timescales \citep{Angles-Alcazar2017,Faucher-Giguere2018,Tacchella2020,Looser2025,McClymont2025}.

While burstiness can be easily measured and quantified in theoretical models, where the evolution of individual galaxies can be traced, observational studies must rely on proxies to assess this phenomenon, as SFHs cannot be measured at arbitrarily short timescales over large lookback times. However, two empirical estimates of SFRs are known to gauge two specific timescales. The \ha\/ emission line traces ionising radiation from massive, short lived stars, therefore probing SFRs on timescales of $\approx5-10$ Myr, while the UV continuum probes SFRs on timescales of $\approx10-50$ Myr \citep{Kennicutt2012}. These tracers have been used to demonstrate that lower-mass galaxies tend to have burstier SFHs \citep{Weisz2012,Guo2015,Emami2019,Boyett2024,Langeroodi2024,Looser2025}. Burstiness has also a direct impact on the SFMS: more bursty SFHs lead to an increase of the SFMS normalisation \citep{Caplar2019, Donnari2019full} and a decrease of the SFMS scatter when longer-timescale SFR indicators are used \citep{Tacchella2020, Wan2025}. Recently, Bayesian hierarchical models have also been explored as a way to infer burstiness from individual and an ensemble of galaxies through spectral energy distribution (SED) modelling \citep{Wan2024,Wan2025,Carvajal-Bohorquez2025}.

Ever since its launch, JWST has given us an unprecedented view of the early Universe \citep{Gardner2023}. To name a few of its advantages that are relevant to this work, it has granted us access to rest-frame optical wavelengths at $z>6$, allowed us to improve SFR indicators through measurements of the Balmer lines (e.g., \ha\/ and \hb\/), and provided us with data to obtain more accurate stellar masses, resulting in updated measurements of the SFMS \citep[][]{Clarke2024,Cole2025,Rinaldi2025}. Importantly, JWST has unveiled an intriguing population of galaxies at $z>10$, which are UV-brighter and more numerous than predicted \citep[e.g., ][]{Naidu2022,Finkelstein2023,Robertson2023,Robertson2024,Carniani2024,Hainline2024,Kokorev2025,Naidu2025,Weibel2025,Whitler2025}. One key avenue to explore in this context, is how bursty SFHs affect -- and possibly explain -- the observed UV luminosity functions at $z>10$ \citep{Shen2023,Sun2023bursty,Kravtsov2024}, specifically, the over-abundance of UV bright galaxies detected with JWST in the past years.

In addition to possibly explaining the puzzling observations at $z>10$, quantifying the burstiness of SFHs has important astrophysical implications at lower redshifts. For example, it can shed light on the sources responsible for ionising the Universe during the EoR. This period of time represents the transition between a dark and neutral IGM to a fully ionised one by $z\sim5-6$ \citep{Keating2020,Bosman2022,Zhu2024}. Star-forming galaxies are believed to have played a crucial role in the ionisation of the Universe, due to the copious amounts of Lyman Continuum radiation (photons with $\lambda<912$ \AA\/, or $E>13.6$ eV) produced by their young massive stars \citep{Hassan2018,Rosdahl2018,Trebitsch2020}. Moreover, in the context of star-forming galaxies as main agents of reionisation, many studies suggest low-mass galaxies with bursty star formation are the dominant source of ionising photons \citep[e.g., ][]{Endsley2023, Yeh2023,Rinaldi2024,Simmonds2024ELGs}. However, it must be noted that recent observations obtained with the JWST have reignited the debate regarding the contribution arising from active galactic nuclei \citep[AGN; see e.g., ][]{Asthana2024,Grazian2024,Maiolino2024, Madau2024}. Therefore, understanding the interplay between stellar mass and star formation as a function of redshift can help unveil the sources of reionisation.

Two important questions that arise from the study of the SFMS are: (1) How universal is the SFMS at early times? and, (2) What processes govern its emergence and evolution? In this work we aim to answer these questions by measuring the shape, scatter, and normalisation of the SFMS for galaxies at $z=3-9$ using JWST NIRCam photometry from the JWST Advanced Deep Extragalactic Survey \citep[JADES; ][]{Eisenstein2023JADES,Bunker2024, D'Eugenio2025}, combined with SED fitting using \texttt{Prospector}. We focus on galaxies at $z=3-9$, where rest-frame optical data are accessible in addition to the UV, enabling robust constraints on stellar masses and SFRs. This is in stark contrast to studies at $z>9$, which typically rely mostly on rest-frame UV photometry and may overestimate the intrinsic scatter in SFR due to larger measurement uncertainties. Moreover, our use of deep JADES photometry (including medium-band data when available) avoids the selection biases and limited completeness inherent in spectroscopic samples, which are affected by strong emission-line flux cuts and complex targeting strategies. The structure of this work is the following. In $\S$~\ref{SEC:data} we discuss the data and sample selection used in this work. We then estimate galaxy properties using the SED fitting code \texttt{Prospector} in $\S$~\ref{SEC:galaxy_properties}, and analyse the stellar mass completeness of our sample. In $\S$~\ref{SEC:SFMS_fitting} we present our best-fit to the SFMS at $3<z<9$, and compare it to estimations from the literature. In $\S$~\ref{SEC:scatter_of_SFMS} we measure the scatter of the SFMS as a function of redshift, stellar mass, and averaging timescales. We then use our results to make conclusions about the burstiness of star formation in the galaxies here studied. We explore the implications of our findings in the context of observations of the early Universe in  $\S$~\ref{SEC:implications}. Finally, we consider the caveats and limitations of our work in $\S$~\ref{SEC:caveats}, ending with concluding remarks in $\S$~\ref{SEC:conclusions}. Throughout this work we assume $\Omega_0=0.315$ and H$_0=67.4$ km s$^{-1}$ Mpc$^{-1}$, following \cite{Planck2020}.

\begin{figure*}	\includegraphics[width=2\columnwidth]{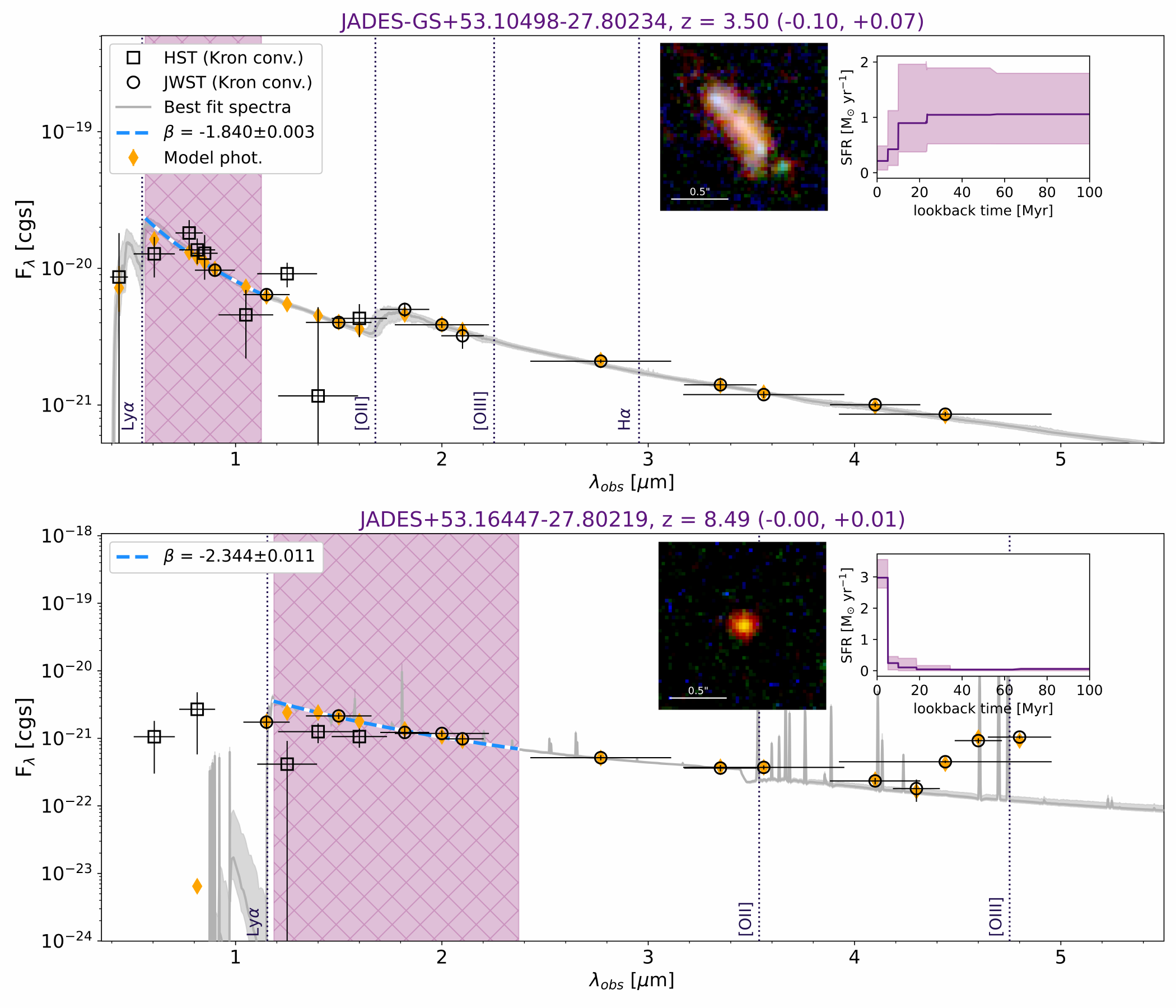}
    \caption{Representative examples of the products of our SED fitting routine with \texttt{Prospector}. The names and redshifts for each galaxy are indicated in the titles. The names are composed by the coordinates of each galaxy rounded to the fifth decimal. The observed photometry is shown as open squares (HST) and circles (JWST NIRCam), while the best-fit photometry is shown as orange diamonds, and the best-fit spectra is shown in grey. We also show the UV continuum slope  \citep[$\beta$, given by F$_\lambda \propto \lambda^{\beta}$][]{Calzetti1994}, obtained by fitting a line to the best-fit spectra in logarithmic space at $\lambda_{\rm{rest-frame}}=1250-2500$\AA\/ (purple hatched region). The vertical dotted lines show the expected wavelength of selected emission lines. Finally, the insets show the star formation history and an RGB image of the selected galaxy. These examples illustrate the power of \texttt{Prospector} to reproduce the observed SED shapes of galaxies even in the absence of strong emission lines. \textsl{Top panel:} galaxy at $z\sim 3.5$ with no evidence of strong emission lines and a declining SFH. \textsl{Bottom panel:} galaxy at $z\sim 8.5$ with strong evidence of \oiii\/$_{\lambda5007}$ and a recent burst in star formation.}  
    \label{fig:SED_examples}
\end{figure*}

\begin{figure}
    \centering \includegraphics[width=\columnwidth]{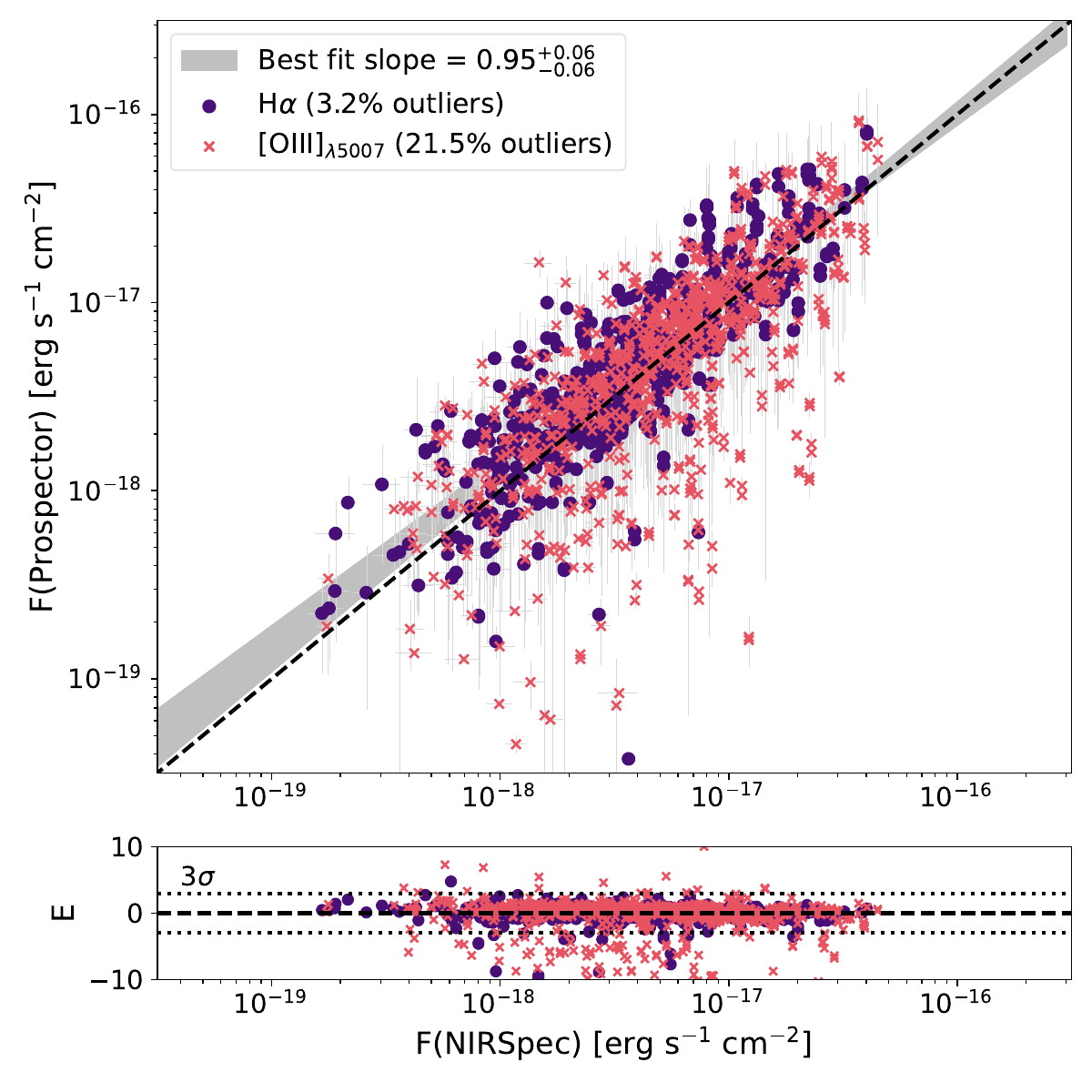}
    \caption{Comparison of \texttt{Prospector}-inferred \ha\ (circles) and \oiii\/$_{\lambda5007}$ (crosses) emission line fluxes to ones measured from JADES using NIRSpec.  
    The bottom panel shows the difference between the inferred and measured values including their respective errors, given by: $\text{E}\equiv(\text{F}_{\rm{Prospector}}-\text{F}_{\rm{NIRSpec}})/(\sqrt{\sigma_{\rm{Prospector}}^{2}+\sigma_{\rm{NIRSpec}}^{2}})$. The dotted lines denote the $3\sigma$ range, measurements outside of these limits are considered outliers. We find an outlier fraction of 3.2\% for \ha\/ and 21.5\% for \oiii\/. Finally, the JADES measurements might be underestimated in bright galaxies due to slit losses. Overall, we find good agreement between our fitted values and the observed ones, as shown by the best fit to the data (shaded area).} 
    \label{fig:flux_comparisons}
\end{figure}

\section{Data and sample selection}
\label{SEC:data}
In this section we briefly describe the observations used in this work, as well as the selection criteria for our sample. 

\subsection{Observations}
We make use of the full JADES \citep{Eisenstein2023JADES,Bunker2024} photometry set in both the GOODS-N and GOODS-S regions \citep{Giavalisco2004}, including the publicly available NIRCam Deep imaging \citep{Rieke2023}, and the JADES Origins Field \citep[JOF; ][]{Eisenstein2023JOF}. When available, we also use photometry from the JWST Extragalactic Medium-band Survey \citep[JEMS; ][]{Williams2023}, and from the First Reionisation Epoch Spectroscopic Complete Survey \cite[FRESCO; ][]{Oesch2023}. 
In addition to the photometric catalogues, we use emission line fluxes derived from spectroscopic observations to compare to fluxes inferred from SED fitting. Specifically, in Section~\ref{sec:prospector_fitting} we compare  our \texttt{Prospector}-inferred fluxes of \ha\/ and \oiii\/$_{\lambda5007}$ to those obtained by  JADES using the JWST micro-shutter assembly (MSA) of NIRSpec \citep{Ferruit2022,Jakobsen2022,Bunker2024,D'Eugenio2025}. 

For the GOODS-S region, we rely on the same photometric catalogues as in \cite{Simmonds2024ELGs} and \cite{Simmonds2024complete}. In this work, however, we also include catalogues produced in the same fashion for the GOODS-N region. Briefly, the source detection and photometry was performed on JEMS NIRCam medium band and JADES NIRCam broad and medium band imaging, using the \texttt{photutils} software package \citep{Bradley2022}. Sources with five or more contiguous pixels of the SNR mosaic with signal $> 3\sigma$ were identified. Furthermore, \texttt{photutils} was used to perform circular aperture photometry with filter-dependent aperture corrections based on model point-spread functions following the method presented in \cite{Ji2024}, as described in \cite{Robertson2024}. In addition to the NIRCam observations, HST images released through the Hubble Legacy Field program \citep{Illingworth2016,Whitaker2019} reductions of GOODS-S and GOODS-N \cite{Giavalisco2004} and Cosmic Assembly Near-infrared Deep Extragalactic Legacy Survey \citep[CANDELS; ][]{Grogin2011,Koekemoer2011} images. In this work, we adopt a Kron aperture on images, convolved to a common resolution, and impose an error floor of 5\% in each band. 

\subsection{Sample selection criteria}
For this work we use \texttt{Prospector}-inferred properties as described in \cite{Simmonds2024complete} for galaxies in GOODS-S, but extended to include galaxies in GOODS-N. Out of the full JADES photometry dataset, we only impose the following conditions: (1) sources must have a signal-to-noise ratio SNR$\geq$3 in the F444W band, (2) a median \texttt{Prospector}-derived redshift in the range $z_{\rm{phot}}=3-9$, and (3) a reduced $\chi^2\leq10$ to remove objects with catastrophically bad fits. Finally, we remove the JWST Mid-Infrared Instrument (MIRI) selected AGN from \cite{Lyu2024} and the broad-line AGN from \cite{Juodzbalis2025}\footnote{We note that these spectroscopically confirmed AGN represent $<$10\% of the entire sample of galaxies for which we have spectra. Moreover, they are mostly composed of low-luminosity AGN, where it is likely that the continuum is dominated by the host galaxy.}. This sample selection criteria yields 29220 galaxies in GOODS-S, and 18936 galaxies in GOODS-N, resulting in a total sample of 48156 galaxies. Out of the total sample, 99\% (91\%) of the galaxies have medium band coverage in at least one (two) photometric band(s).

\section{Estimation of Galaxy properties}
\label{SEC:galaxy_properties}
This work uses galaxy properties derived from the SED modelling code \texttt{Prospector}, specifically, we infer stellar masses, SFRs across different timescales, and UV magnitudes. In this section, we outline the SED modelling procedure, how SFRs are estimated, and the stellar mass completeness of our sample. SED modelling offers a more robust 
estimation of SFRs compared to direct tracers like UV and H$\alpha$, as it incorporates a galaxy's full SED, accounting for systematic errors that may arise from variations in dust attenuation (incl. variations in the dust attenuation law), complex SFHs, and nebular continuum and line emission. While UV and H$\alpha$ tracers are affected by factors such as dust attenuation and metallicity (e.g. through their impact on conversion factors), they often provide limited insight into the full SFH, especially since they trace specific timescales. For example, H$\alpha$ is insensitive to timescales larger than a few tens of million years, as it is a tracer of instantaneous star formation. In contrast, SED modelling with sufficient spectral coverage, resolution, and sensitivity enables a self-consistent analysis that mitigates systematic biases and yields a more accurate view of star formation across timescales, particularly at high redshift, by properly accounting for parameter degeneracies and uncertainties.

\subsection{SED fitting with Prospector}
\label{sec:prospector_fitting}
In this work, we use the SED fitting code \texttt{Prospector} \citep{Johnson2019, Johnson2021} to derive galaxy properties.
The fitting routine used to fit our entire sample is described in \cite{Simmonds2024complete}. 
Briefly, we use photometry from the full JADES photometric catalogues as input for \texttt{Prospector}, including data from the UV to the IR from HST and JWST NIRCam \citep[see figure 1 from ][]{Simmonds2024complete}. We adopt the same Kron convolved aperture to extract HST, JADES and JEMS photometry, and floor the photometric errors to 5\% to account for systematic errors not accounted for by our data reduction pipelines. We use photometric redshifts (z$_{\rm{phot}}$) inferred with the SED template fitting code \texttt{EAzY} \citep{Brammer2008} as priors in our SED fitting routine. As shown in \cite{Rieke2023} and \cite{Hainline2024}, z$_{\rm{phot}}$ can be accurately determined when a rich photometry set is used. 
The dust attenuation and stellar population properties are varied following \cite{Tacchella2022prosp}. Importantly, we use a two component dust model \citep{Charlot2000, Conroy2009}, which accounts for the differential dust attenuation in young ($<10$ Myr) stars and nebular emission lines by using different optical depths and a variable dust index \citep{Kriek2013}. We adopt a Chabrier \citep{Chabrier2003} initial mass function (IMF) with mass cutoffs of 0.1 and 100 M$_{\odot}$, and allow the stellar metallicity to explore a range between 0.01 and 1 Z$_{\odot}$. The continuum and emission properties of the SEDs are provided by the Flexible Stellar Population Synthesis (FSPS) code \citep{Conroy2010}, based on \texttt{Cloudy} \citep[v13.03; ][]{Ferland2013} using MESA Isochrones \& Stellar Tracks \citep[MIST; ][]{Choi2016, Dotter2016}, and the MILES stellar library \citep{Vazdekis2015}. The UV extension of the latter is based on the Basel Stellar Library \citep[BaSeL; ][]{Lastennet2002}. The use of this \texttt{Cloudy} grid results in a limitation of the maximum permitted ionisation parameter of log$\langle$U$\rangle_{\rm{max}}$=-1.0, which might not be enough to reproduce the properties of some high-redshift galaxies \citep[e.g., ][]{Cameron2023}. We set a flexible IGM model based on a scaling of the Madau model \citep{Madau1995}, where the scaling is left as a free parameter. Finally, we use a non-parametric SFH \citep[continuity SFH; ][]{Leja2019}. In this model, the SFH is described as eight different SFR bins. The minimum bin is set to a look-back time of $1-5$ Myr and $5-10$ Myr, while the rest are divided equally in logarithmic space depending on the redshift. The ratios and amplitudes of the bins are allowed to vary following a Student's t-distribution with a width of 0.3. Importantly, this prior allows (but does not impose) a somewhat bursty SFH when the data favour it. 

To showcase the products of our SED fitting process, in Figure~\ref{fig:SED_examples} we present two galaxies with significantly different SFHs at the limits of our redshift range. These are representative examples that highlight the power of \texttt{Prospector} when a comprehensive photometric set is used. In particular, they demonstrate how \texttt{Prospector} is able to fit the observed SED shapes even in the absence of strong emission lines \citep[see also figure A.2 of][]{Simmonds2024complete}.  
Moreover, we find that \texttt{Prospector} can reproduce emission line fluxes reasonably well for our sample based on photometry alone. In Figure~\ref{fig:flux_comparisons} we compare  \ha\/ and \oiii\/$_{\lambda5007}$ fluxes derived by \texttt{Prospector} to a subsample of galaxies with JADES NIRSpec observations (782 \ha\/ and 887 \oiii\/ flux measurements with SNR$\geq5$). 
Overall, we find good agreement between our SED-inferred fluxes and those measured by spectroscopic observations, with values scattering around the 1:1 line and outlier fractions of 3.2\% and 21.5\% for \ha\/ and \oiii\/, respectively. Therefore, we confirm that our SED fitting routine can reproduce emission lines to a reasonable extent. 

\subsection{Stellar mass completeness}
\label{sec:completeness}
As in \cite{Simmonds2024complete}, we estimate the stellar mass completeness of our sample following the method described in Section 5.2 of \cite{Pozzetti2010}. To summarise, for every redshift a minimum mass (M$_{\rm{min}}$) is defined, above which potentially all types of galaxies can be observed (i.e., not only those who have had a recent burst in star formation). To calculate M$_{\rm{min}}$, the limiting stellar mass (M$_{\rm{lim}}$) must be estimated. M$_{\rm{lim}}$ represents the stellar mass a galaxy would have if its apparent magnitude (m) were equal to the limiting magnitude of the survey (m$_{\rm{lim}}$). Importantly, M$_{\rm{lim}}$ depends on the stellar mass-to-light ratio, a quantity that can be affected by bursty star formation (i.e., by galaxies being up-scattered due to a recent burst in star formation). In this work m$_{\rm{lim}}$ is determined by the F444W band since it has the highest effective wavelength, making it the best tracer of stellar mass available in our data, although we caution that at $z>5$ it is susceptible to strong emission line biases. For each galaxy, M$_{\rm{lim}}$ is given by:
\begin{equation}
    \log(\text{M}_{\rm{lim}}) = \log(\text{M}_{\star})+0.4(\text{m}-\text{m}_{\rm{lim}}),
\end{equation}
where M$_{\star}$ is stellar mass in units of solar masses. We divide our sample into three fields depending on their F444W exposure time: medium ($\text{T}_{\rm{exp}} < 25$ ks), deep ($25~\text{ks} \leq \text{T}_{\rm{exp}} < 65$ ks) and ultra-deep ($\text{T}_{\rm{exp}} \geq 65$ ks), with $5\sigma$ flux depths of 6, 4.5, and 2.65 nJy, respectively. In order to compute M$_{\rm{min}}$ for each field and at each redshift bin, we select the faintest 20 per cent of galaxies for each exposure time. This choice takes into account the colour-luminosity relation and therefore includes only galaxies with a
typical $M/L$ close to the magnitude limit.

\begin{figure}	\includegraphics[width=\columnwidth]{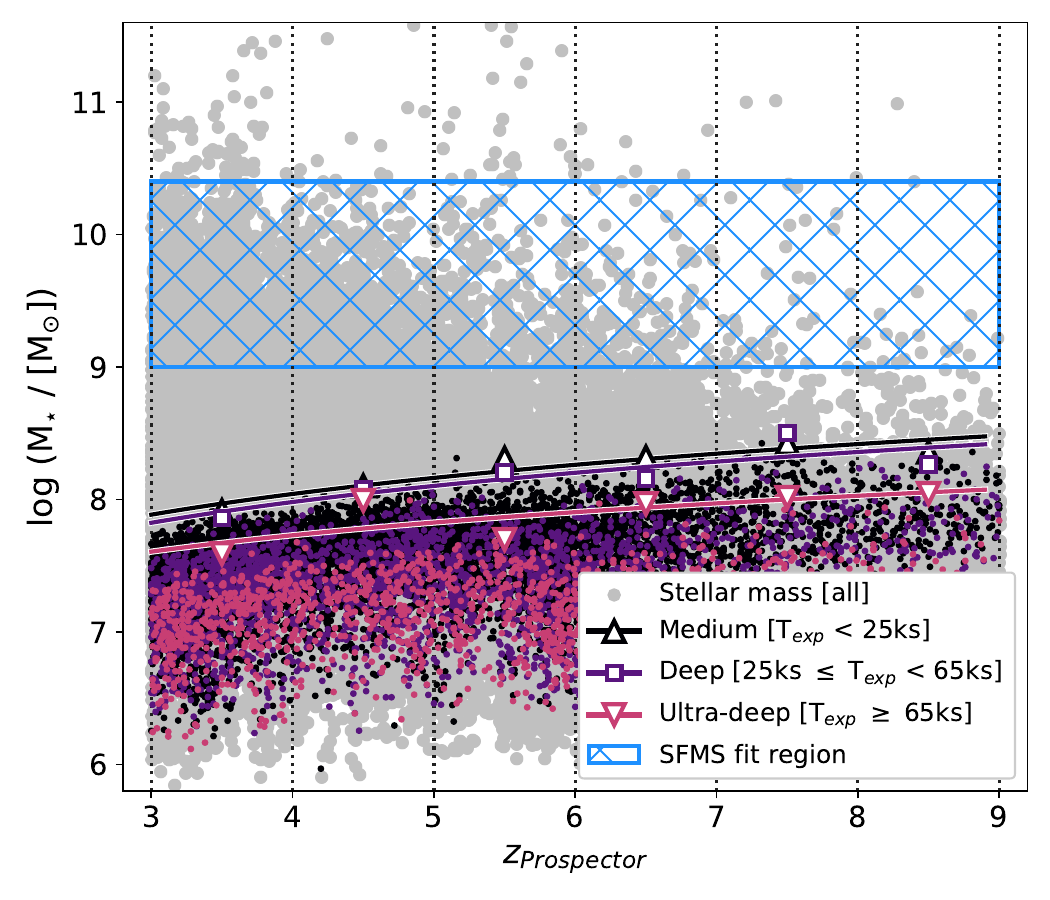}
    \caption{Stellar mass completeness of the sample, divided into three fields depending on exposure time (\texp\/). The larger grey circles show the stellar mass of each galaxy as a function of redshift, while the smaller coloured circles show the limiting mass for the faintest 20\%. The solid curves denote the 100\% completeness at each redshift and for each depth, as indicated in the legend, following the prescription of \citet{Pozzetti2010}. We find our sample is stellar mass complete down to log(M$_{\star}$/[M$_{\odot}$]) $\approx 8.1$ on average, with small variations depending on the depth of the field. The dotted vertical lines mark the redshift bins used in the mass completeness estimation. The blue hatched area shows the region used to fit the star-forming main sequence (i.e. $9.0\leq$log(M$_{\star}$/M$_{\odot}$)$\leq10.3$ and $3\leq z\leq9$.)}
    \label{fig:mass_completeness}
\end{figure}

In \cite{Simmonds2024complete} we focused on galaxies above which 90 per cent of the selected M$_{\rm{lim}}$ lie, however, in this work we place a stricter limit of 100 per cent, because even a small incompleteness percentage can significantly bias estimates of average SFRs \citep{Leja2022}. The results are shown in Figure~\ref{fig:mass_completeness}, where the stellar mass as a function of redshift is shown for our full sample (grey circles), and the M$_{\rm{lim}}$ of the faintest 20 per cent of galaxies per T$_{\rm{exp}}$ are shown as coloured points. We find, on average, that our sample is stellar mass complete down to log(M$_{\star}$/M$_{\odot}$)$\approx$8.1. We note that the method used here to establish the stellar mass completeness limits of our sample relies on the assumption of a constant mass-to-light ratio. If SFHs are bursty, it means that SFRs change on short timescales, which results in a non-constant mass-to-light ratio in the ionising (and to a lesser extent in the non-ionising) continuum \citep{Dominguez2015}. Thus, if the galaxies in our sample have bursty SFHs, then the limits shown in Figure~\ref{fig:mass_completeness} should be taken as lower limits, as will be discussed further in Section~\ref{sec:implications_mass}.

In the remainder of this work, we focus on galaxies with masses higher than the stellar mass completeness limit derived above of log(M$_{\star}$/M$_{\odot}$)$\approx8.1$, yielding a sample of 18321 galaxies ($\sim38$\% of the total sample). However, we also present a more conservative stellar mass completeness limit motivated by the behaviour of the sSFR as a function of stellar mass (see Section~\ref{SEC:SFMS_fitting}), with log(M$_{\star}$/M$_{\odot}$)=9.0-10.3, shown as a blue hatched region in Figure~\ref{fig:mass_completeness}, containing 2989 galaxies.

\begin{figure*}	\includegraphics[width=\columnwidth]{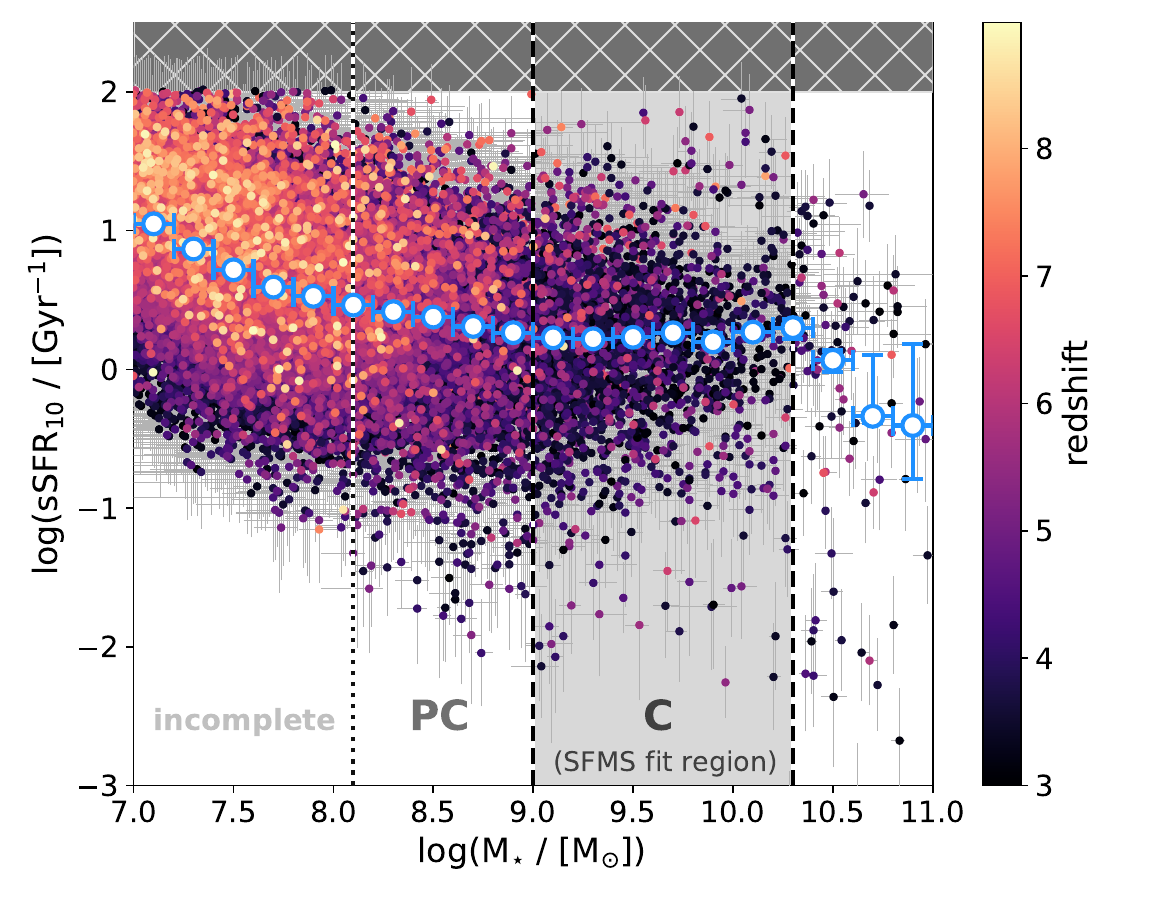}
\includegraphics[width=0.9\columnwidth]{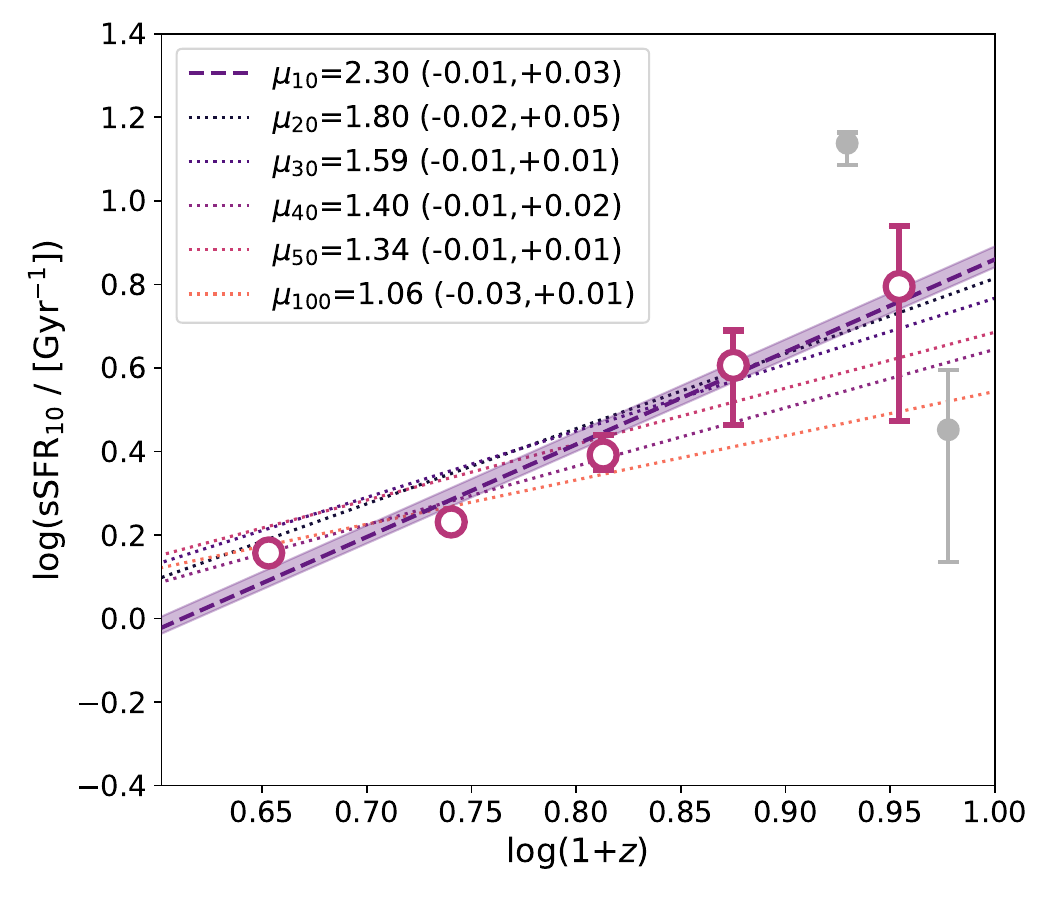}
    \caption{Stellar mass range and redshift evolution used to fit the star forming main sequence. \textsl{Left panel:} specific star formation rate (sSFR$_{10}$) using the star formation averaged over the past 10 Myr (SFR$_{10}$), as a function of stellar mass for all galaxies in our sample with log(M$_{\star}$/M$_{\odot}$)$\geq 7.0$, colour-coded by redshift. The white circles with blue edges show the medians per stellar mass bins (0.2 width in logarithmic space). The vertical lines show different stellar mass completeness limits, delimiting three regions. The upper limit of the region labelled "incomplete" is given by the mean stellar mass completeness estimated in Figure~\ref{fig:mass_completeness} (log(M$_{\star}$/M$_{\odot}$)$\approx 8.1$). We note that the median sSFRs increase steadily below this limit. This trend is consistent with an increasing prevalence of rising SFHs at lower stellar masses, but it is also expected in biased samples that are incomplete in stellar mass. Interestingly, this behaviour continues (although less steeply) up to log(M$_{\star}$/M$_{\odot}$)$=9.0$, which we dub partially complete ("PC"). The shaded area shows galaxies with $9.0\leq$log(M$_{\star}$/M$_{\odot}$)$\leq10.3$ (also shown as a blue hatched area in Figure~\ref{fig:mass_completeness}), we note that the sSFR flattens in this mass range, and thus we define this region as being complete ("C"), and use all galaxies in it to fit the star forming main sequence (i.e., without imposing a SFR range). The upper limit of the complete region is motivated by the turnover of the SFMS at log(M$_{\star}$/M$_{\odot}$)$=10.3$. Finally,  the hatched horizontal area shows the upper limit allowed given the timescale used. \textsl{Right panel:} median sSFRs for galaxies with $9.0\leq$log(M$_{\star}$/M$_{\odot}$)$\leq9.5$ as a function of redshift, the mass bin was selected because it is populated at all redshift bins. To increase numbers, we merge the two highest redshift bins ($7\leq z\leq 9$). We fit a line to all medians to obtain the redshift evolution, $\mu_{10}$, which we use as input when fitting the SFMS (see Equation~\ref{eq:MS}). The redshift evolution for all the remaining timescales analysed in this work are shown as dotted lines.} 
    \label{fig:mass_completeness_true}
\end{figure*}

\begin{figure*}
    \centering	\includegraphics[width=2\columnwidth,trim={5cm 0 5cm 0}]{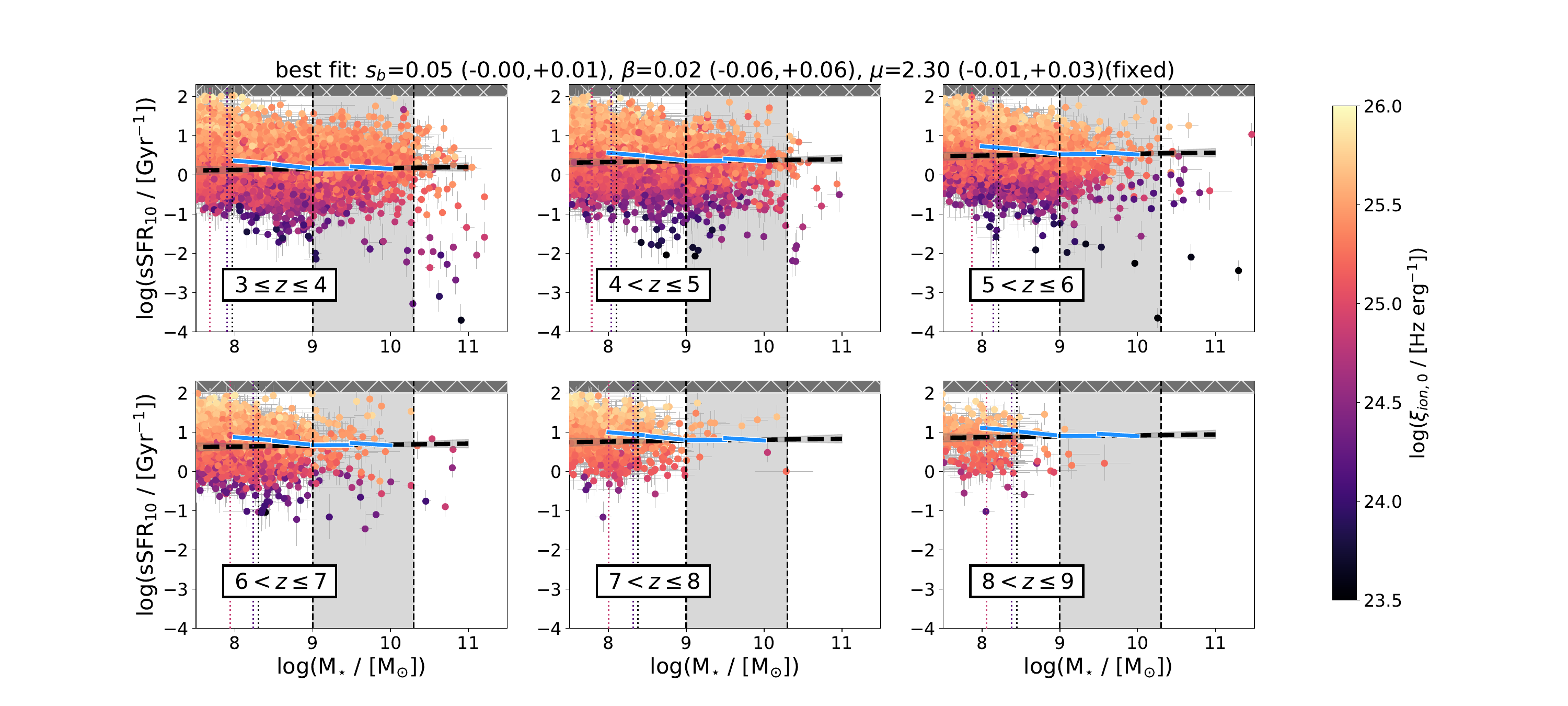}
    \caption{Specific star formation rate, sSFR$_{10}$, as a function of stellar mass, colour-coded by their ionising photon production efficiency (\xionnofesc\/), and divided by redshift bins. The thick black dashed lines show the best fit to the star-forming main sequence given in Equation~\ref{eq:MS}, for galaxies with stellar masses in the range \(9.0 \leq \log(\text{M}_{\star}/\text{M}_{\odot}) \leq 10.3\) (indicated by the grey shaded region bounded by vertical dashed lines), and are extrapolated to all stellar masses in our sample. To improve the robustness of the fit at high redshift, the two highest redshift bins were grouped together when deriving the best-fit relations, though they are shown separately in the figure for consistency with the redshift binning used throughout the paper. The blue filled lines show piece-wise fits to the same equation in mass bins of stellar masses log(M$_{\star}$/M$_{\odot}$)=8.0-10.0 in steps of 0.5 in logarithmic space. The grey hatched region shows the upper limit in sSFR, given by $\frac{1}{\text{t}_{\rm{average}}}$. For reference, we include the stellar mass completeness limits shown in Figure~\ref{fig:mass_completeness}: from left to right, the vertical dotted lines show the ultra-deep, deep, and medium exposure times.  We note that if we fit the SFMS using these limits instead (log(M$_{\star}$/M$_{\odot}$)$\gtrapprox 8.1$), the slope of the fit becomes negative, as indicated by the discrepancy between the piece-wise fit and the total fit in that stellar mass regime.}
\label{fig:MS_fit_10_complete}
\end{figure*}

\begin{figure*}
    \centering	\includegraphics[width=2\columnwidth]{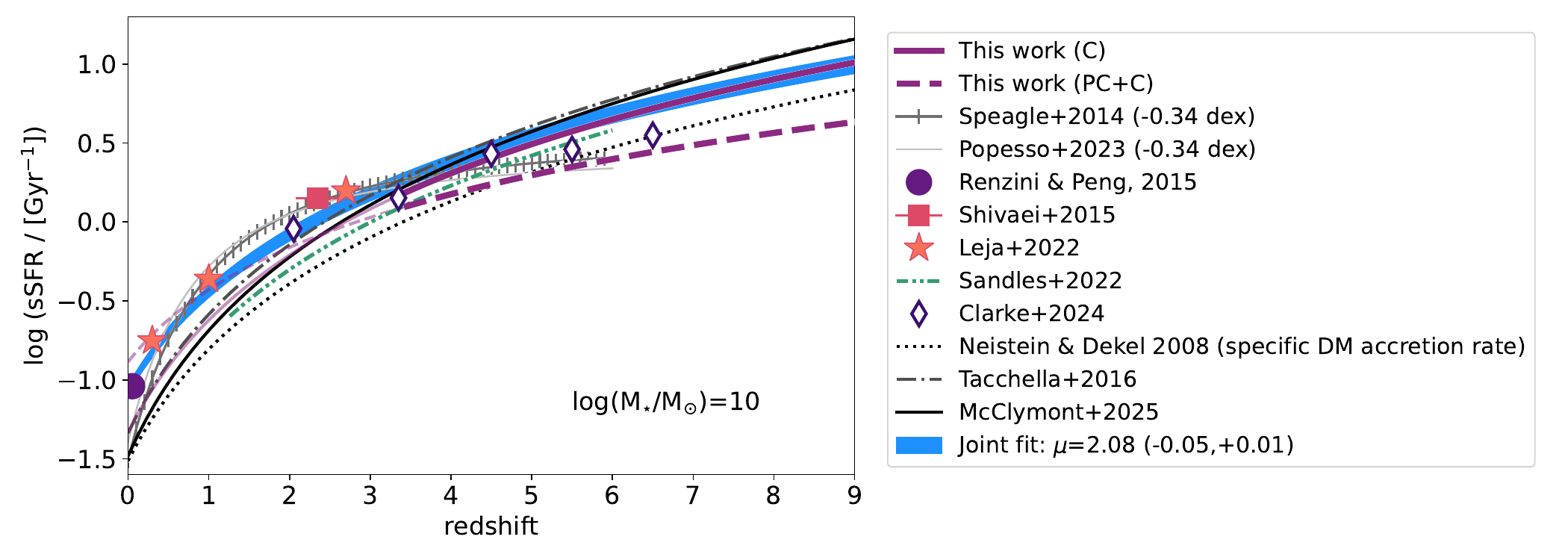}
    \caption{Comparison of the SFMS derived in this work to selected ones from the literature for recent star formation (either averaged over the past 10 Myr or derived from \ha\/ fluxes). All markers and curves are shown for a fixed stellar mass of 10$^{10}$M$_{\odot}$. The purple filled and dashed curves show the the SFMS fit to the complete ("C") and partially complete plus complete ("PC+C") samples, extrapolated to $z=0$. The markers show the results for studies performed at redshift bins up to $z=7$ \citep[][, as indicated in the legend]{Renzini2015, Shivaei2015,Leja2022,Clarke2024}. The grey curve shows the SFMS from \citet{Popesso2023}, while the light grey line with vertical markers shows the \citet{Speagle2014} relation, both shifted  by -0.34 dex to account for conversions in SFRs, as in \citet{Shapley2023}. The green dash-dotted curve shows the parametrisation provided by \citet{Sandles2022}. The grey dash-dotted curve shows the results from \citet{Tacchella2016}, where only the normalisation s$_{\rm{b}}$ was left as a free parameter, and the black filled line shows the SFMS based on the \textsc{THESAN-ZOOM} simulations from \citet{McClymont2025}. The discrepancies at $z<3$ can be partially explained due to differences in how the stellar masses were inferred, and how complete the samples are in stellar mass, but mainly arise due to the sample selections and conversions of \ha\/ luminosities to SFRs. The black dotted curve shows the expected specific mass accretion ratio of dark matter halos from \citet{Neistein2008}, using the M$_{\star}$-M$_{\rm{halo}}$ conversion from \citet{Tacchella2018}. It is important to note that using a different conversion would result in slightly shifting this curve, but that the redshift evolution of the SFMS found in this work is remarkably similar to the specific mass accretion ratio of dark matter halos.  
    Finally, a joint fit to the shifted \citet{Speagle2014} up to $z=3$ with our fit to the complete sample ($z=3-9$) is shown as a blue shaded area. Including the lower redshift results from literature in the fit results in a slightly shallower evolution with redshift ($\mu\approx 2.08$ instead of $2.30$).}
    \label{fig:literature_comparison}
\end{figure*}

\begin{figure}
    \centering	\includegraphics[width=1\columnwidth]{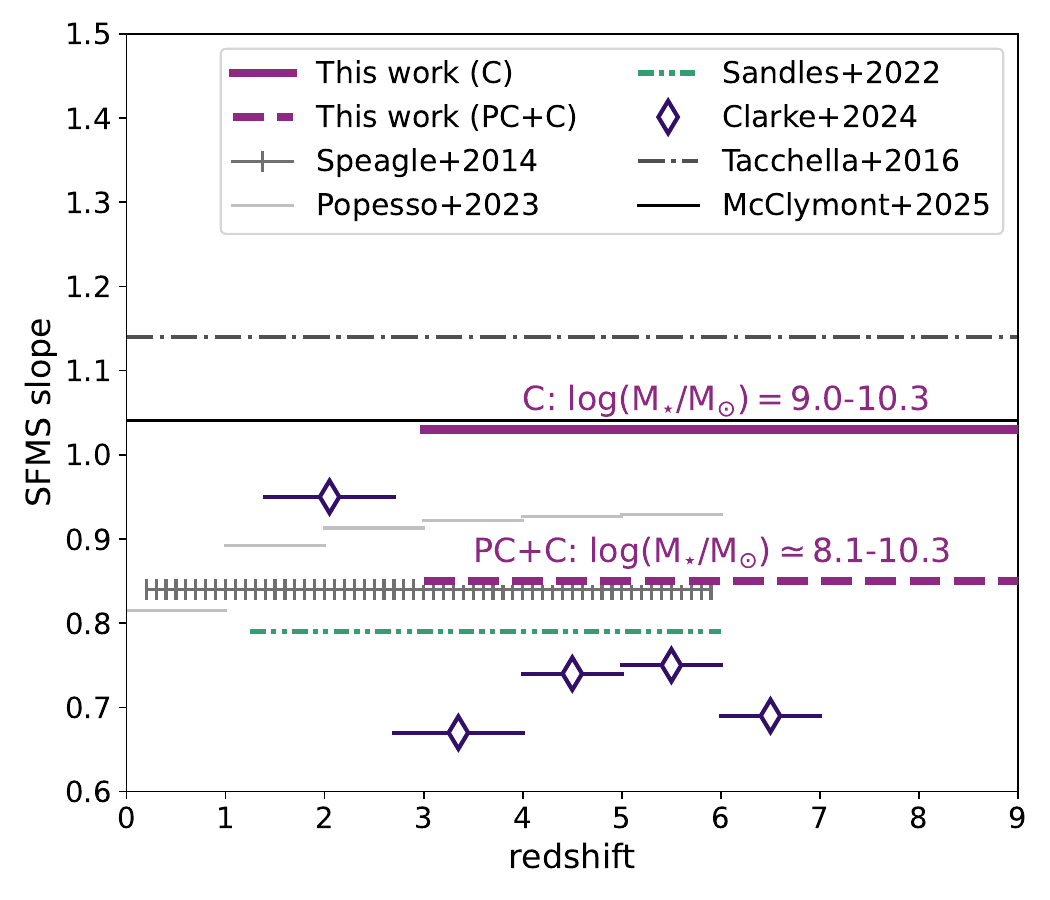}
    \caption{Comparison of the SFMS slope, defined as the slope of the logarithmic relation between SFR$_{10}$ and stellar mass ($\mathrm{SFR}\propto \mathrm{M}_{\star}^{\alpha}$, where $\alpha=1+\beta$ with $\beta$ from Eq.~\ref{eq:MS}), as a function of redshift for a selection of works shown in Figure~\ref{fig:literature_comparison}. As can be seen, the slope when considering the complete ("C") sample approaches 1, while for the partially complete plus complete sample ("PC+C") the slope is shallower. This discrepancy is potentially driven by stellar mass completeness.}
    \label{fig:literature_comparison_slopes}
\end{figure}

\subsection{Star formation rates} 
We have shown in Figure~\ref{fig:flux_comparisons} that \texttt{Prospector} can retrieve emission line fluxes reasonably well. However, we decide against using SFR tracers derived from these lines \citep[e.g., from \ha\/; ][Kramarenko et al. in prep.]{Kennicutt1998,Pflamm-Altenburg2009,Kennicutt2012,Tacchella2022} or from the UV luminosity due to their dependence on dust corrections, metallicity and the parameter space on which the relations have been calibrated, both introducing additional uncertainties \citep{Rosa-Gonzalez2002, Kennicutt2012, Smit2016, Curtis-Lake2021, Shapley2023}\footnote{We note that SFR$_{10}$ and SFR$_{\rm{H}\alpha}$ largely agree for galaxies in our sample, both when \texttt{Prospector}-inferred and NIRSpec measured H$\alpha$ fluxes are used (see Appendix~\ref{app:SFR10_SFRHa}).}. Moreover, SFR calibrations based on UV and IR luminosities have been shown to be systematically higher than SFRs inferred through SED fitting at a given stellar mass \citep{Bauer2011}, especially if the galaxies are quiescent \citep[more than $\sim1$ dex; ][]{Hayward2014,Utomo2014,Leja2022}. This could be an indication of obscured star formation, which can be difficult to detect in the rest-frame UV and optical wavelength range \citep[i.e., "outshining"; ][]{Papovich2001,Conroy2013,Whitaker2017,Papovich2023,Sun2024,Williams2024,Herard-Demanche2025,Reddy2025}. Quantifying this effect on our sample is beyond the scope (and limitations) of this work, however, future surveys with JWST MIRI are a promising avenue to explore this problem further \citep{Helton2025SED}. In addition to dust corrections and conversions between luminosity and SFRs, this discrepancy can be explained by older stellar populations contributing to the UV and IR regions of galaxy SEDs, and is more significant in the absence of young stellar populations (i.e., a recent star burst). Therefore, in this work, we rely on the SFRs provided by \texttt{Prospector} at different averaging timescales, as they offer a more comprehensive estimation by incorporating a galaxy's full SED, including dust attenuation, SFHs, and nebular emission. This approach reduces the systematic biases associated with direct tracers like UV and H$\alpha$, and allows for a self-consistent analysis across multiple timescales, which is particularly important when studying galaxies at high redshifts where such tracers may be affected by additional complexities. 

Specifically, we use the SFRs averaged over the past 10, 20, 30, 40, 50, and 100 Myr (indicated by the subscript of the SFR used). These SFR can be estimated largely independent given the time bins in our SFH modelling, e.g., any covariance between these SFRs can be attributed physical processes that drive star formation. These values are obtained by integrating the SFHs and dividing them by the stellar mass formed at each timescale. The self-consistency provided by the \texttt{Prospector} modelling imposes a natural maximum in sSFR given by $\frac{1}{\text{t}_{\rm{average}}}$. Studying different timescales of star formation for a given galaxy informs us about the burstiness of its SFH. In particular, the ratio between the star formation averaged over two different timescales (i.e. SFR$_{10}$/SFR$_{50}$ or SFR$_{10}$/SFR$_{100}$) contains information about the fluctuations in star formation over time \citep{Weisz2012,Guo2016,Caplar2019, Clarke2024, Cole2025}.

\begin{table*}
	\centering
	\caption{Best fit values to SFMS given in Equation~\ref{eq:MS} for $9.0\leq$log(M$_{\star}$/M$_{\odot}$)$\leq10.3$ and $3\leq z\leq9$. \textsl{Column 1:} averaging timescale of the SFR in units of Myr. \textsl{Column 2:} normalisation in units of Gyr$^{-1}$. \textsl{Column 3:} stellar mass dependence. \textsl{Column 4:} redshift evolution. \textsl{Columns 5,6:} normalisation of the SFMS, $\eta(z)=\mathrm{s}_{\rm{b}}(z)\times(1+z)^{\mu}$, at a fixed stellar mass of log(M$_{\star}$/M$_{\odot}$)=10.0, calculated at $z=3$ and $z=9$, respectively.}
	\label{tab:best_fit_values}
	\begin{tabular}{cccccc} 
		\hline
		t$_{\rm{average}}$ [Myr] & s$_{\rm{b}}$ [Gyr$^{-1}$] & $\beta$ & $\mu$ & $\eta (z=3)$ [Gyr$^{-1}$] & $\eta (z=9)$ [Gyr$^{-1}$] \\
		\hline
		10 & $0.05_{-0.01}^{+0.01}$ & $0.02_{-0.06}^{+0.08}$ & $2.30_{-0.01}^{+0.03}$ & $1.21_{-0.24}^{+0.24}$ & $9.98_{-2.00}^{+2.00}$ \\[0.15cm]
        20 & $0.12_{-0.01}^{+0.01}$ & $0.03_{-0.06}^{+0.05}$ & $1.80_{-0.02}^{+0.05}$ & $1.46_{-0.12}^{+0.13}$ & $7.57_{-0.63}^{+0.63}$ \\[0.15cm]
        30 & $0.17_{-0.01}^{+0.02}$ & $0.04_{-0.05}^{+0.05}$ & $1.59_{-0.01}^{+0.01}$ & $1.54_{-0.09}^{+0.18}$ & $6.61_{-0.39}^{+0.78}$ \\[0.15cm]
        40 & $0.21_{-0.02}^{+0.02}$ & $0.03_{-0.04}^{+0.05}$ & $1.40_{-0.01}^{+0.02}$ & $1.46_{-0.14}^{+0.14}$ & $5.27_{-0.50}^{+0.50}$ \\[0.15cm]
        50 & $0.23_{-0.02}^{+0.02}$ & $0.02_{-0.05}^{+0.05}$ & $1.34_{-0.01}^{+0.01}$ & $1.47_{-0.13}^{+0.13}$ & $5.03_{-0.44}^{+0.44}$ \\[0.15cm]
        100 & $0.30_{-0.02}^{+0.02}$ & $0.00_{-0.04}^{+0.04}$ & $1.06_{-0.03}^{+0.01}$ & $1.30_{-0.09}^{+0.09}$ & $3.44_{-0.23}^{+0.23}$ \\
		\hline
	\end{tabular}
\end{table*}

\section{Fitting the star forming main sequence}
\label{SEC:SFMS_fitting}

The SFMS describes the relation between SFR (or sSFR) and stellar mass, and has been extensively studied up to $z\sim6$ \citep[][]{Brinchmann2004,Daddi2007, Elbaz2007,  Salim2007, Chen2009, Pannella2009, Santini2009, Magdis2010, Oliver2010,Elbaz2011,Karim2011,Rodighiero2011,Shim2011,Lee2012,Reddy2012,Salmi2012,Whitaker2012,Zahid2012,Kashino2013,Moustakas2013,Sobral2014,Speagle2014,Steinhardt2014,Whitaker2014,Lee2015,Schreiber2015,Tasca2015,Shivaei2015,Erfanianfar2016,Santini2017,Belfiore2018,Pearson2018,Popesso2019a,Popesso2019b,Leslie2020,Thorne2021,Daddi2022,Leja2022,Rinaldi2022,Sandles2022,Baker2023,Popesso2023,Clarke2024,Davies2025P1,Davies2025P2,Cole2025,Rinaldi2025}.
Comparing SFMS measurements from the literature is non-trivial because of the use of different selection techniques and wavelengths \citep[see, e.g., ][]{Katsianis2016}. Therefore, studies such as \cite{Popesso2023} and \cite{Speagle2014} have focused on converting all measurements to a common calibration, finding a consensus in the shape and normalisation of the SFMS. 
Basically, there are three main factors that must be considered when studying the SFMS: (1) the estimation of the SFRs, obtained via dust-corrected emission line luminosities and/or SED fitting, (2) the sample selection, and (3) the stellar mass completeness of the sample. The latter is of utmost importance because the process of fitting the SFMS is highly sensitive to the mass range used (see Section~\ref{sec:SFMS_bf_params}). Thanks to surveys such as JADES, we are now able to construct large samples of galaxies complete in stellar mass up to $z\sim9$.

The SFMS has been shown to evolve with redshift, both in observations \citep[e.g.,][]{Speagle2014, Popesso2023} and simulations \citep[e.g., ][]{Tacchella2016,D'Silva2023,McClymont2025}. A convenient way to take this evolution into account when fitting the SFMS is to fit for the sSFR, adopting the following form:
\begin{equation}
\label{eq:MS}
    \text{sSFR}_{\rm{MS}}(\text{M}_{\star},z)[\text{Gyr}^{-1}] = \text{s}_{\rm{b}}\times \bigg(\frac{\text{M}_{\star}}{10^{10}\text{M}_{\odot}}\bigg)^{\beta}\times(1+z)^{\mu},
\end{equation}
as in \cite{Tacchella2016}. In this equation, the sSFR of the main sequence is determined by the redshift evolution ($\mu$) of the SFMS, a stellar mass dependence ($\beta$, not to be confused with the UV continuum slope), and an overall normalisation (s$_{\rm{b}}$, in units of Gyr$^{-1}$). We assume in this work no curvature of the SFMS masses at high stellar masses, since we focus on galaxies with stellar masses below $10^{11}~M_{\odot}$.

The same functional form as described by Eq.~\ref{eq:MS} can be used to describe the specific mass accretion rate (sMAR) of dark matter halos. In this case, \cite{Dekel2013} show that $\mu_{\rm h}$ tends to $5/2$ when a $\Lambda$CDM cosmology in the Einstein-deSitter regime is adopted. The halos mass dependence, $\beta_{\rm h}$, is expected to contribute only in a minor way, with $\beta_{\rm h}\approx 0.14$ \citep{Neistein2008}. The normalisation of the sMAR, as measured from simulations, is $s_{\rm h}=0.03~\mathrm{Gyr}^{-1}$ \citep{Dekel2013}. While the sSFRs of SFMS galaxies do not need to track the sMAR of their halos, the normalisation of the SFMS has been shown to increase with look-back time as $\mathrm{sSFR}\propto(1+z)^{\sim2.2}$ \citep{Brinchmann2004, Daddi2007, Elbaz2007, Noeske2007, Whitaker2012, Speagle2014, Schreiber2015, Boogaard2018,Caplar2019, Donnari2019,Sandles2022}, which is very close to the expected value from the sMAR \citep{Bouche2010,Lilly2013, Tacchella2013, Tacconi2013, Rodriguez-Puebla2016, Tacchella2018}.  

In order to fit Equation~\ref{eq:MS} to our data, we carefully selected the parameter space by following two steps, which are now briefly explained and are summarised in Figure~\ref{fig:mass_completeness_true}.

\begin{figure*}
    \centering	\includegraphics[width=1.8\columnwidth,trim={5cm 1cm 5cm 0}]{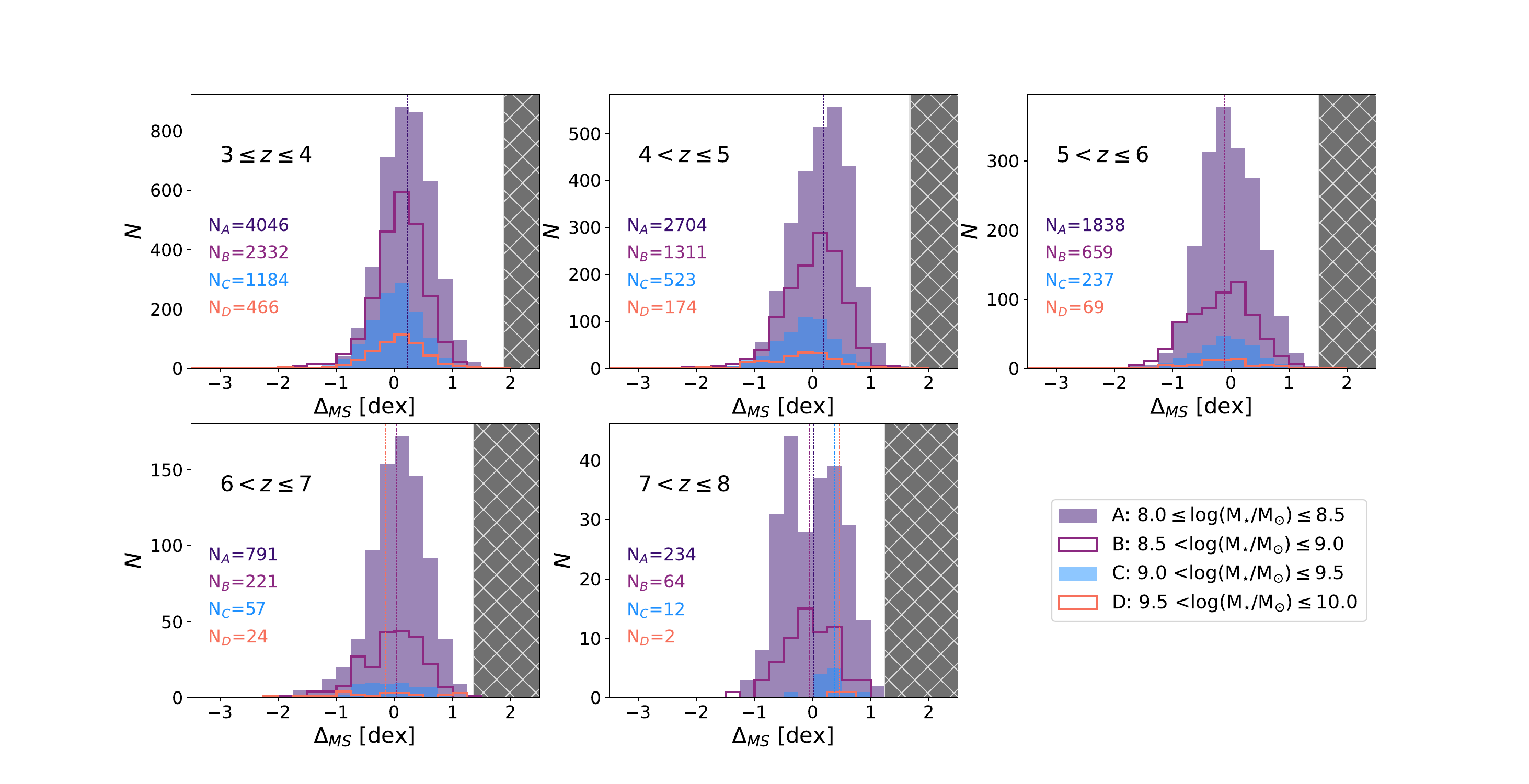}
    \caption{Histograms showing the distance to the SFMS, $\Delta_{\rm{MS}}$, using the star formation averaged in the past 10 Myr (shown as thick black dashed lines in Figure~\ref{fig:MS_fit_10_complete}), as function of redshift bin up to $z=8$. We define four stellar mass bins, at log(M$_{\star}$/M$_{\odot}$)=8.0-10.0, as indicated in the legend. For each panel, the number of galaxies in each mass bin is shown, as well as the median of each distribution. The grey hatched region shows the maximum permitted $\Delta_{\rm{MS}}$ at each redshift, for log(M$_{\star}$/M$_{\odot}$)=8.0. Although composed of very few galaxies ($<100$ in total), the histograms tend to have a tail towards the lower values and thus cannot be described appropriately by a Gaussian distribution.}
    \label{fig:scatter_10_linear}
\end{figure*}

\begin{figure*}
    \centering	\includegraphics[width=2\columnwidth]{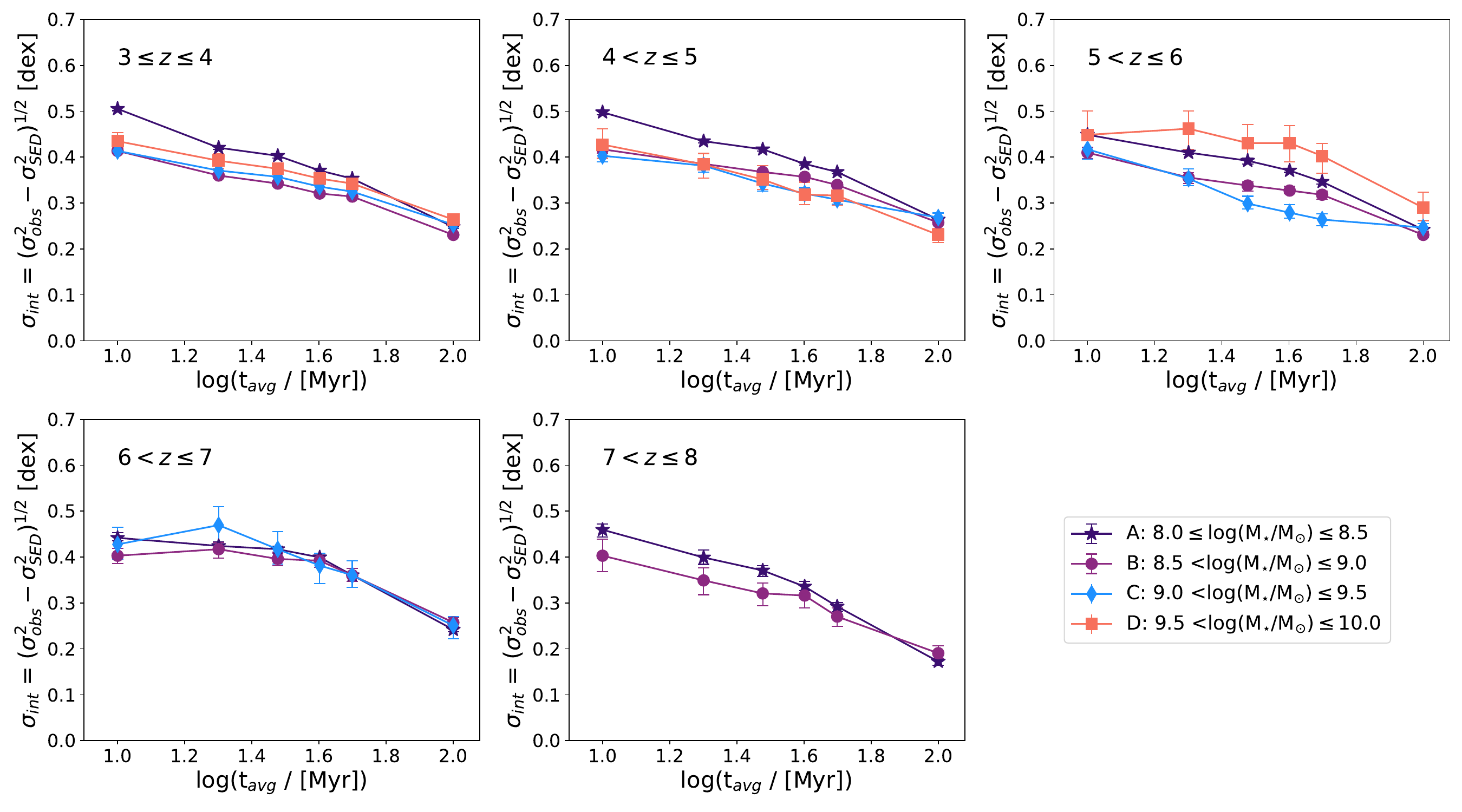}
    \caption{Intrinsic scatter of the SFMS, $\sigma_{\rm{int}}$, as a function of redshift and stellar mass, as indicated in the legend, where $\sigma_{\rm{int}} = \sqrt{\sigma_{\rm{obs}}^2-\sigma_{\rm{SED}}^2}$. Only bins with at least 30 galaxies are shown. The intrinsic scatter decreases as averaging timescales increase for every mass and redshift bin studied, from $\sigma_{\rm{int}}\approx 0.4-0.5$ at 10 Myr down to $\sigma_{\rm{int}}\approx 0.2$ at 100 Myr. This behaviour is expected when star formation is bursty, as averaging over longer timescales smooths out episodic bursts and (mini-)quenching events, reducing the scatter in the SFRs.} 
    \label{fig:scatter_zevolution}
\end{figure*}

\subsection{Mass range used in SFMS fitting}
The slope of the SFMS is highly sensitive to the stellar mass range being used. This range should reflect the stellar mass completeness of the sample. In Section~\ref{sec:completeness}, we derive this limit following \cite{Pozzetti2010}, and mention that it might be an underestimation of the stellar mass completeness if star formation is bursty, since it relies on a constant mass-to-light ratio. As a conservative approach, we decide to select the range of stellar mass we use in the SFMS by studying the behaviour of the sSFR as a function of stellar mass. In the left panel of Figure~\ref{fig:mass_completeness_true} we show the relation between sSFR and stellar mass when using the shortest timescale presented in this work, SFR$_{10}$ (comparable to SFRs derived with \ha\/). We find that the median sSFRs steadily decrease with stellar mass up to log(M$_{\star}$/M$_{\odot}$)$\sim 8.1$, in a region that we define as "incomplete". We note that the median sSFRs increase steadily below this limit. This trend is consistent with an increasing prevalence of rising SFHs at lower stellar masses, but it is also expected in biased samples that are incomplete in stellar mass. Since this trend becomes more pronounced with increasing incompleteness (e.g., at lower stellar masses) and is not reproduced in numerical simulations by increasing SFHs \citep{McClymont2025}, we interpret it as primarily driven by sample incompleteness.

This behaviour continues up to log(M$_{\star}$/M$_{\odot}$)$= 9.0$, albeit with a shallower slope, suggesting that this region still suffers from stellar mass completeness issues. This could be due to a combination of \texttt{Prospector} limitations when inferring SFRs at the lower mass end, of low-mass galaxies having preferentially increasing SFHs, and of SFHs being burstier than expected at this mass range, and is discussed further in Section~\ref{sec:implications_mass}. We dub this region as "partially complete" ("PC"). Finally, we find that the sSFRs flatten at $9.0\leq$log(M$_{\star}$/M$_{\odot}$)$\leq 10.3$, where our upper mass limit agrees with the turnover observed by \cite{Leja2022} \citep[but see also ][ where a turnover is observed at M$_{\star}$$\sim10^{10}$ M$_{\odot}$ at $z<1.3$]{Lee2015}. We label this region as "complete" ("C") and use it to fit the SFMS hereafter.

\subsection{Redshift evolution used in SFMS fitting}
In this work we divide our sample into equally spaced redshift bins between $z=3-9$, and note that as redshift increases, the number of galaxies decreases. As a result, if we fit Equation~\ref{eq:MS} directly, leaving $\mu$, $\beta$, and $s_{\rm{b}}$ as free parameters, our SFMS fits are dominated by the bulk of the galaxies at $z\leq5$. Therefore, we constrain the redshift evolution of our sample separately (and for each SFR timescale) and fix it as median and boundaries when fitting the SFMS. The right panel of Figure~\ref{fig:mass_completeness_true} shows this exercise for SFR$_{10}$, where we have merged the two highest redshift bins ($7\leq z\leq 9$), and fit a line in log space to the median sSFRs for galaxies with $9.0\leq$log(M$_{\star}$/M$_{\odot}$)$\leq9.5$, as a function of redshift. We choose to use this stellar mass bin since it contains galaxies at all redshifts here studied. For this averaging timescale of 10 Myr we find $\mu=2.30^{+0.03}_{-0.01}$, a value that is similar to others found in the literature \citep[e.g., ][]{Sandles2022,Popesso2023,McClymont2025}. We then extrapolate this redshift dependence of the SFMS to all stellar masses in our sample, and repeat this exercise with all the timescales analysed in this work (shown as dotted lines and given in Table~\ref{tab:best_fit_values}). 

\subsection{SFMS best-fit parameters}
\label{sec:SFMS_bf_params}
Figure~\ref{fig:MS_fit_10_complete} shows the results of fitting Equation~\ref{eq:MS} to our data (see also Table~\ref{tab:best_fit_values}), after following the previous steps, for SFRs averaged over the past 10 Myr and per redshift bin. The thick black dashed lines show the best fits to the SFMS (using galaxies in the ``complete'' region, shown as a shaded grey area). For reference, we show the stellar mass completeness limits derived following \cite{Pozzetti2010} as vertical lines. To show how sensitive the SFMS fits are to the stellar mass limits, we show piece-wise fits for stellar masses of log(M$_{\star}$/M$_{\odot}$)$=8.0-10.0$ in logarithmic steps of 0.5 (blue filled lines). It can be seen that there is a discrepancy between the piece-wise fits and the general fit at masses below log(M$_{\star}$/M$_{\odot}$)$=9.0$, which is a result of this region being only partially complete in stellar mass. The individual points are colour-coded by their ionising photon production efficiency assuming an escape fraction of zero, \xionnofesc\/, estimated by \texttt{Prospector} \citep[see e.g., ][]{Simmonds2023, Simmonds2024ELGs}. This quantity is essentially the ratio between recent (averaged over the past 10 Myr) star formation, and the one sustained over the past 100 Myr \citep{Simmonds2024ELGs}. Although this ratio directly probes the recent SFH, its variance for an ensemble of galaxies measures short-term star formation variability \citep{Caplar2019}. As expected, the distance to the SFMS is related to \xionnofesc\/, where the galaxies that are most efficient producers of ionising radiation lie above the relation.

\subsection{The normalisation of the SFMS} 
The best-fitting parameters for all adopted SFR averaging timescales are listed in Table~\ref{tab:best_fit_values}. We find that the normalisation \(\text{s}_{\rm{b}}\) increases with averaging timescale, while the redshift dependence $\mu$ decreases with averaging timescale. While \(\text{s}_{\rm{b}}\) quantifies the normalisation of the SFMS at a given redshift as per Equation~\ref{eq:MS}, we must also account for the explicit redshift evolution through the \((1+z)\)-term (and given by $\mu$) to compare absolute normalisation values across timescales (see Figures~\ref{fig:mass_completeness_true} and \ref{fig:literature_comparison}). At \(z=3\) and log(M$_{\star}$/M$_{\odot}$)$=10$, we measure a SFMS normalisation, which we define as $\eta(z)=\mathrm{s}_{\rm{b}}(z)\times(1+z)^{\mu}$, of 1.2~Gyr$^{-1}$ for \(t_{\rm{average}} = 10\)~Myr and 1.5~Gyr$^{-1}$ for \(t_{\rm{average}} = 50\)~Myr. At \(z=9\), we find a SFMS normalisation of 10.0~Gyr$^{-1}$ for \(t_{\rm{average}} = 10\)~Myr and 5.0~Gyr$^{-1}$ for \(t_{\rm{average}} = 50\)~Myr (see Table~\ref{tab:best_fit_values}). 

The interpretation of this trend is not straightforward, as it reflects the interplay of two competing effects: bursty star formation and the overall rise of SFHs. As shown by \citet[][see also \citealt{Donnari2019full,Iyer2020}]{Caplar2019}, the SFMS normalisation depends on both the stochasticity of star formation and the averaging timescale of the SFR measurement. Short-timescale indicators capture rapid fluctuations, including low-SFR episodes, whereas long-timescale indicators smooth over these variations, leading to higher inferred average SFRs and an elevated normalisation. An effective tool to understand the power contained in SFR fluctuations at a specific timescale is the power spectral density (PSD). On short timescales (\(<10\) Myr), galaxy growth is governed by internal processes such as GMC formation, star formation, and stellar feedback \citep{Leitherer1999,Tan2000,Tasker2011,Faucher-Giguere2018,Benincasa2020}. On intermediate timescales (\(<100\) Myr), large-scale feedback and gas cycling dominate \citep{Semenov2017,Shin2023}, while on longer timescales, external drivers such as mergers become important \citep{Robertson2006,Puskas2025}. Crucially, the dependence of SFMS normalisation on the averaging timescale, \(\text{t}_{\rm{average}}\), encodes information about the stochastic nature of galaxy SFHs, e.g. its PSD \citep{Caplar2019}. The higher SFMS normalisation on  the 50~Myr relative to the 10~Myr averaging timescale at $z=3$ is consistent with a break timescale of \(\tau_{\rm{break}} \approx 50\)~Myr and a PSD slope \(\alpha=2\), in agreement with the expectations of bursty star formation and the predictions of \citet[][see their figure~5]{Caplar2019}. However, towards higher redshifts, the trend reverts and the normalisation is higher for shorter averaging timescales.

\begin{figure*}
    \centering	\includegraphics[width=2\columnwidth]{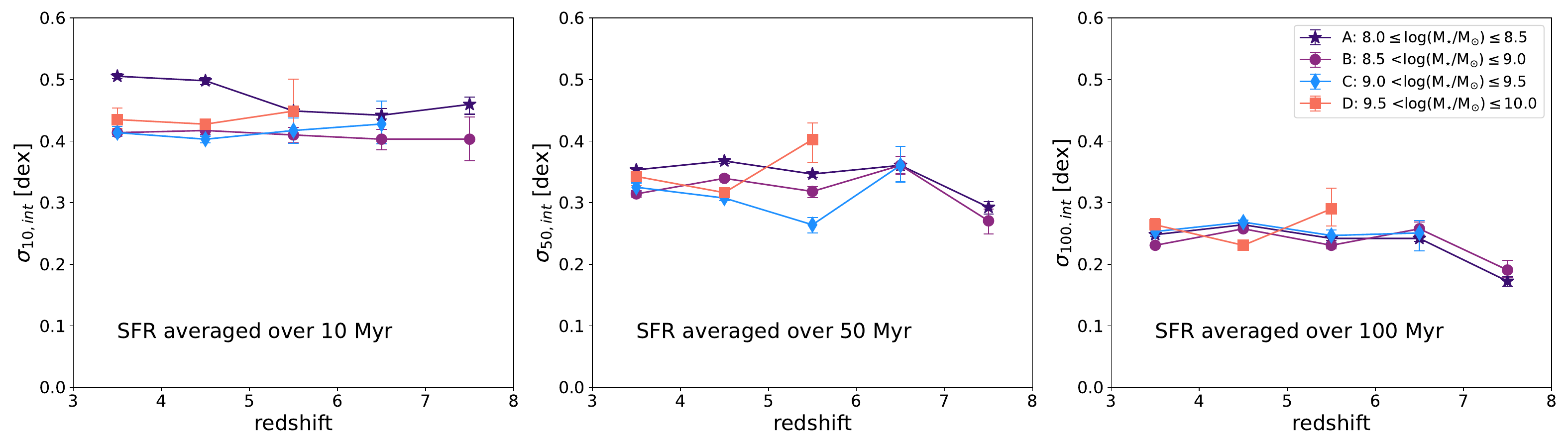}
    \caption{Intrinsic scatter of the SFMS, $\sigma_{\rm{int}}$, as a function of redshift, for three selected averaging timescales, from left to right: 10, 50, and 100 Myr. Only bins containing at least 30 galaxies are shown. As expected, $\sigma_{\rm{int}}$ decreases as timescales increase. Interestingly, we observe no strong dependence of $\sigma_{\rm{int}}$ on redshift.}    \label{fig:scatter_timescales_vs_z}
\end{figure*}

\begin{figure}
    \centering	\includegraphics[width=1\columnwidth]{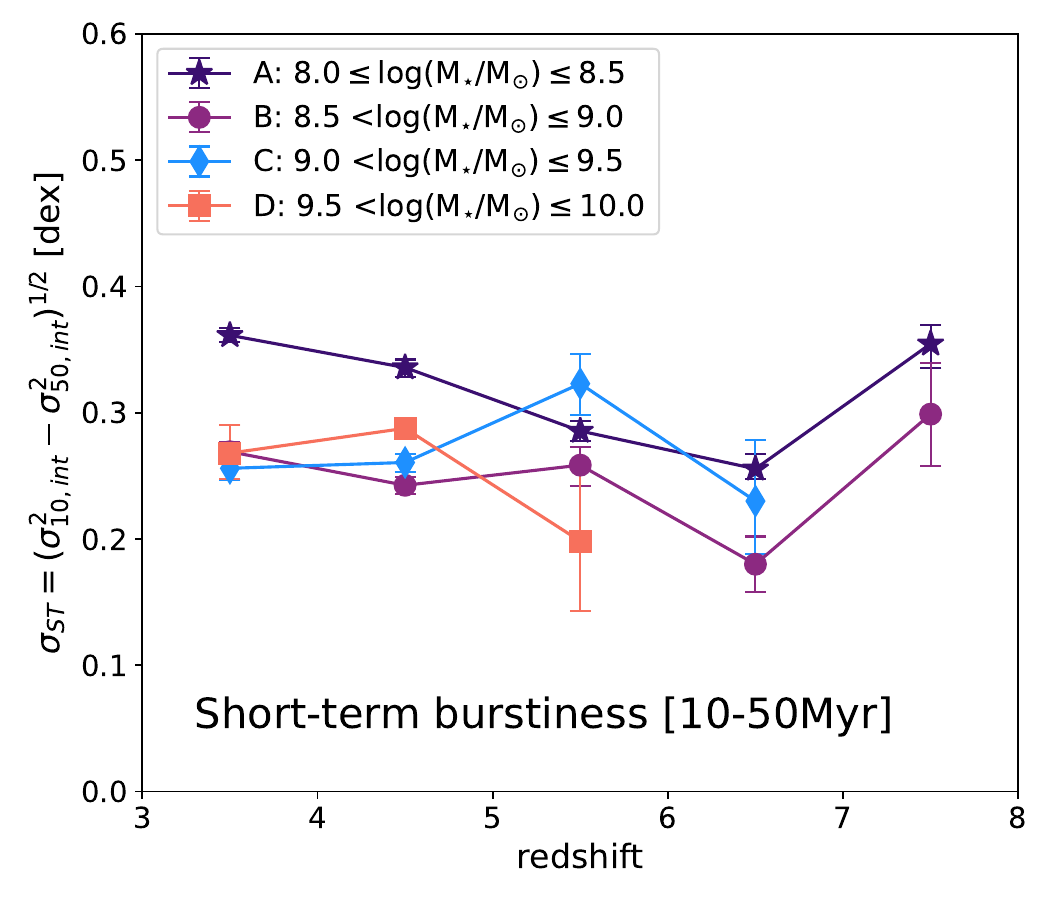}
    \caption{Short-term burstiness, $\sigma_{\rm{ST}}$, defined as the difference between the intrinsic scatter of the SFMS between a look-back time of 10 and 50 Myr, as a function of redshift. Our findings suggest SFHs are bursty for all of the stellar mass bins here considered. We draw attention to the lowest mass bin ("A"), which contains the largest amount of galaxies, and has indications of being slightly burstier. In this mass bin, we see that $\sigma_{\rm{ST}}$ decreases between $z=3$ and $z=6$, and increases at $z>7$. This can be interpreted as an indication of two different mechanisms dominating the short-term scatter of the SFMS: environmental effects at $z>8$, and internal processes at $z<8$. We note that the trends persist if we look at the difference between look-back times of 10 and 100 Myr instead.}
    \label{fig:scatter_short_timescales}
\end{figure}

A second factor affecting the interpretation of the SFMS normalisation is the presence of rising SFHs. For galaxies with rising SFHs, the SFR averaged over shorter timescales is systematically higher than when averaged over longer timescales. Consequently, if most galaxies follow rising SFHs, the SFMS normalisation is expected to be higher when measured on short timescales. We expect, in particular towards higher redshift, the fraction of galaxies with increasing SFHs to increase \citep[e.g.,][]{Tacchella2018}. Indeed, for galaxies in our sample with $8.1\lesssim$log(M$_{\star}$/M$_{\odot}$)$\leq10.3$, the fraction of galaxies with SFR$_{10}$/SFR$_{100}>2$ increases from 0.31 to 0.55, from $z=3$ to $z=9$. The elevated SFMS normalisation measured on shorter timescales at \(z > 6\) is therefore primarily driven by the prevalence of rising SFHs in early galaxies. While this complicates the use of normalisation alone as a diagnostic of star-formation variability, the scatter around the SFMS provides a more direct probe of burstiness (see Section~\ref{subsec:scatter_estimation}). In future work (Bevins et al., in prep.), we will build on this analysis by forward-modelling the scatter as a function of \(t_{\rm average}\), enabling tighter constraints on the PSD and the temporal coherence of star formation in the high-redshift Universe.

\subsection{Comparison of our SFMS to literature}
We now compare our obtained SFMS (Eq.~\ref{eq:MS} and Table~\ref{tab:best_fit_values}, using sSFR$_{10}$) with other studies in the literature, both at the same redshift and lower redshifts. In the comparison, we focus on the normalisation of the SFMS (Figure~\ref{fig:literature_comparison}) and the slope that describes the stellar mass dependence (Figure~\ref{fig:literature_comparison_slopes}). Specifically, we compare our SFMS to $z<7$ studies by \cite{Renzini2015}, \cite{Shivaei2015}, \cite{Leja2022}, \cite{Sandles2022}, and \cite{Clarke2024}, as well as the results from \cite{Speagle2014} and \cite{Popesso2023}, which are both based on compilations of SFR measurements\footnote{Both the \cite{Speagle2014} and the \cite{Popesso2023} relations have been shifted by a factor of -0.34 dex, as suggested by \cite{Shapley2023}, to account for SFR conversions.} up to $z=6$. \citet{Renzini2015} and \citet{Shivaei2015} derived SFRs from H$\alpha$ luminosities using Balmer decrement–based dust corrections for local SDSS galaxies and MOSFIRE observations for galaxies at $z\sim2$, respectively. \cite{Leja2022} derived their SFMS from SED fitting with the \texttt{Prospector} code, focusing on galaxies at $0.5<z<3$, while \cite{Sandles2022} used SED fitting with \texttt{BEAGLE} \citep{Chevallard2016} to derive the SFMS at $1.25<z<6$. Finally, we include the recent measurements by \cite{Clarke2024}, based on $\sim150$ galaxies at $1.4<z\leq7$, who used Balmer line luminosity, SED fitting, and UV luminosity to derive SFRs.

Figure~\ref{fig:literature_comparison} compares the normalisation of the SFMS from this work using sSFR$_{10}$ at fixed stellar mass of log(M$_{\star}$/M$_{\odot}$)$=10.0$ (for both the "complete" and "complete+partially complete" samples, extrapolated to $z=0$) to other selected results from the literature. We provide an updated SFMS prescription (indicated with the thick blue line) from $z=0$ to $z=9$, which slightly flattens the redshift evolution, particularly at $z>3$ (with $\mu\approx 2.08$). We find overall good agreement between our and the literature values, with offsets of less than 0.3 dex,, which can be attributed to varying stellar mass and SFR measurement methods. In particular, \cite{Sandles2022} find a redshift evolution of $\mu=2.40\pm0.18$, which is consistent with our work. We also compare our results to simulations by \cite{McClymont2025} and \cite{Tacchella2016}, and find that our SFMS matches the predictions from these models. Moreover, we find that the normalisation of the SFMS follows the specific dark matter accretion rate from \cite{Neistein2008} \citep[see also ][]{Lilly2013}, in agreement with the conclusions from \cite{Sandles2022}.

Figure~\ref{fig:literature_comparison_slopes} shows the slope of the SFMS as a function of redshift from our work and estimates from the literature. We define SFMS slope as the slope of the logarithmic relation between SFR$_{10}$ and stellar mass (e.g., $\mathrm{SFR}\propto \mathrm{M}_{\star}^{\alpha}$, where $\alpha=1+\beta$ with $\beta$ from Eq.~\ref{eq:MS}). Since the SFMS slope is significantly more affected by sample selection and completeness than the SFMS normalisation, it is not surprising that we find a larger scatter in values from the literature. Specifically, we note that many observational studies derive a slope to the M$_{\star}$-SFR relation that is less than 1 \citep[e.g., $\alpha\approx0.7-0.9$; ][]{Speagle2014, Sandles2022,Popesso2023,Clarke2024}, which would be negative in M$_{\star}$-sSFR (e.g., $\beta<0$). We define our conservative completeness cut based on where the slope in M$_{\star}$-sSFR starts to become negative. Indeed, mass incompleteness leads to a shallower derived slope of M$_{\star}$-SFR: discrepancies in the SFMS shape at $z<6$ can be explained by stellar mass incompleteness and sample biases, with lower mass galaxies being underrepresented \citep[as demonstrated in][]{McClymont2025}. We demonstrate the difference in the derived slopes if we push to $10^8$ M$_{\odot}$ into the "partially complete" regime, which yields a slope of 0.85 (compared to 1.03 in the complete regime).

\section{Scatter of the SFMS}
\label{SEC:scatter_of_SFMS}
Beyond the average relation between SFR and stellar mass, the scatter around the SFMS encodes valuable information about the diversity of galaxy growth histories and the stochasticity of mass assembly across cosmic time. Previous studies have found intrinsic scatter values that slightly vary with redshift and stellar mass, with $\sigma_{\rm{int}}\leq0.5$ dex \citep{Behroozi2013,Kurczynski2016,Sandles2022}, while others report little-to-no dependence on stellar mass \citep[$\sigma_{\rm{int}}\sim 0.2-0.3$ dex; ][]{Elbaz2007,Noeske2007,Schreiber2015}. We stress that these studies mostly focused on massive galaxies, above $10^9~M_{\odot}$. Burstiness of star formation tends to be associated with low stellar masses \citep{Weisz2012,Guo2015} due to the relative importance of stellar feedback \citep{Stinson2007,Dome2024}. In low-mass galaxies, this has been attributed to shallow gravitational potentials being unable to retain their ISM after the onset of stellar (e.g., SNe) feedback \citep{Hopkins2023}, as well as to the inherently stochastic nature of star formation driven by the finite lifetimes of GMCs and their discrete sampling \citep{Tacchella2020}. The increase of burstiness with redshift has been attributed to increasingly stochastic IGM inflow \citep{McClymont2025}, but also may be due to the shorter dynamical timescales of high-redshift galaxies \citep{Faucher-Giguere2018,Tacchella2020}. In support of the relation between bursty star formation and scatter of the SFMS, several works have found the scatter to be higher for low-mass galaxies, both in observations \citep{Kauffmann2014,Weisz2014,Santini2017,Boogaard2018,Atek2022,Asada2024,Cole2025} and simulations \citep{Shen2014,Dominguez2015,Iyer2020,McClymont2025}. We note that this mass dependence is not always observed. For example, \cite{Kurczynski2016}, focusing on $z\sim0.5-3$, do not find a significant dependence of the intrinsic scatter on stellar mass. However, they cannot rule out this behaviour since the smallest averaging timescale they study is $\sim 100$ Myr, which would smooth out bursty star formation on shorter timescales. Interestingly, some studies argue for an increase of the scatter as a function of stellar mass for galaxies with log(M$_{\star}$/M$_{\odot})=10.0-11.5$ \citep{Guo2013,Popesso2019a,Popesso2019b,Sherman2021,Cole2025}, but we caution that low-number statistics and that quenching of star formation in massive galaxies could contribute to boosting the scatter \citep{Whitaker2014,Lee2015,Tomczak2016,Baker2025quench}. This mass range is outside of the scope of this work, since we do not have the statistics to study this stellar mass bin at all redshifts.  

\subsection{Estimating the scatter of the SFMS}
\label{subsec:scatter_estimation}
To estimate the scatter of the SFMS as a function of redshift and stellar mass, we define four stellar mass bins as follows:
\begin{itemize}
    \item A: $8.0\leq$log(M$_{\star}$/M$_{\odot}$)$\leq 8.5$ 
    \item B: $8.5 <$log(M$_{\star}$/M$_{\odot}$)$\leq 9.0$ 
    \item C: $9.0 <$log(M$_{\star}$/M$_{\odot}$)$\leq 9.5$ 
    \item D: $9.5 <$log(M$_{\star}$/M$_{\odot}$)$\leq 10.0$ 
\end{itemize}
Since we do not have any galaxies in bin D at $z>8$, we only focus on redshift bins up to $z=8$. We calculate the logarithmic distance to the SFMS fits, $\Delta_{\rm{MS}}$, for each timescale (given by inserting the best-fit parameters in Table~\ref{tab:best_fit_values} into Equation~\ref{eq:MS}), as a function of redshift and stellar mass bins. The resulting histograms for sSFR$_{10}$ are shown in Figure~\ref{fig:scatter_10_linear}, where the hatched grey area shows the timescale-dependent maximum SFR limit ($\mathrm{SFR_{\rm max}}=\text{M}_{\star}/\text{t}_{\rm average}$). The number of galaxies in each stellar mass bin is shown per redshift bin. As can be seen, the distributions tend to have a tail towards lower $\Delta_{\rm{MS}}$ values. Moreover, we are likely missing galaxies in this region \citep{Feldmann2017}. Although the number of galaxies that compose these tails is small ($<100$ galaxies in total), they have the effect of making the distributions non-Gaussian. Therefore, in order to obtain the observed scatter of the SFMS, we mirror these distributions around zero \citep[which has been shown to be a good estimation by ][]{McClymont2025} and define the observed scatter, $\sigma_{\rm{obs}}$, as the standard deviation of the now Gaussian distributions (see Appendix~\ref{app:mirror}). 

After obtaining $\sigma_{\rm{obs}}$, we estimate the intrinsic scatter, $\sigma_{\rm{int}}$, by subtracting the uncertainties on stellar mass and SFR from SED modelling in quadrature, and obtain errors by bootstrapping the distributions 100 times each. The SED errors, which are obtained by averaging the uncertainties inferred by \texttt{Prospector} per stellar mass and redshift bin, vary between $\approx 0.1-0.3$ dex: they decrease as a function of averaging timescale and increase as a function of redshift. We note that these errors do not account for systematics such as outshining. In Figure~\ref{fig:scatter_zevolution} we show the intrinsic scatter obtained after following the steps described above, for bins where we have at least 30 galaxies. As expected, the intrinsic scatter decreases as a function of averaging timescale from $\sigma_{\rm{int}}\approx0.4-0.5$ at 10 Myr to $\sigma_{\rm{int}}\approx0.2$ at 100 Myr, for all bins studied, with slight variations depending on the stellar mass analysed, in agreement with previous studies \cite[e.g., ][]{Elbaz2007,Cole2025}. Figure~\ref{fig:scatter_timescales_vs_z} shows $\sigma_{\rm{int}}$ for three different look-back times as a function of redshift, from left to right: 10, 50, and 100 Myr for bins containing at least 30 galaxies. As expected, the scatter decreases as the averaging timescales increase, from $\sim0.4$ to $\sim 0.2$. Our measurements provide a robust constraints on the intrinsic scatter of the SFMS across a wide stellar mass range at $z=3-8$, based on homogeneous, high-quality photometry. We find no strong trend in the scatter with redshift, indicating a relatively stable level of variability in star formation over this epoch. However, we do observe a weak mass dependence, with slightly larger scatter at lower stellar masses at shorter timescales. These findings are broadly consistent with theoretical expectations \citep[e.g.,][]{McClymont2025}.

\subsection{Constraint on star-formation burstiness}
We estimate short-term burstiness, $\sigma_{\rm{ST}}$, by measuring the difference in the intrinsic scatter of the SFMS between averaging timescales of 10 and 50~Myr, thereby isolating the contribution from rapid, short-term processes on timescales of 10~Myr to the ones on timescales of 50~Myr:
\begin{equation}
    \sigma_{\rm{ST}}=\sqrt{\sigma_{\rm{10,int}}^2-\sigma_{\rm{50,int}}^2},
\end{equation}
as in \cite{McClymont2025}. 

Figure~\ref{fig:scatter_short_timescales} shows the short-term burstiness, $\sigma_{\rm{ST}}$, as a function of redshift for the four stellar mass bins defined above. This quantity represents the first robust measurement of burstiness imprinted onto SFMS scatter, specifically isolating the contribution from rapid fluctuations in star formation. We find that $\sigma_{\rm{ST}} \approx 0.2-0.4$ dex across all redshifts and mass bins, indicating that short-term variability is a significant and persistent feature of galaxy growth during EoR ($z=3-8$). While there is no strong trend with stellar mass, we observe that the lowest-mass bin ($8.0\leq$log(M$_{\star}$/M$_{\odot}$)$\leq 8.5$) shows systematically higher short-term burstiness, with a dip at $z = 6-7$. The significance and origin of this feature remain uncertain due to observational limitations, and we caution that its interpretation is speculative.

Interestingly, a similar dip in $\sigma_{\rm{ST}}$ at $z \sim 6-7$ has been reported in the \textsc{THESAN-ZOOM} simulations \citep{McClymont2025}, where it arises despite a decrease in the overall scatter $\sigma_{\rm{int,10}}$ with redshift. Thanks to the ability to disentangle different sources of variability, these simulations show that short-term burstiness is primarily driven by internal processes (e.g., feedback, stochastic star formation), but that externally driven variability -- particularly fluctuations in inflow from the circumgalactic medium -- becomes increasingly important at higher redshift. This supports the interpretation that while internal processes dominate short-term variability, the slight increase in $\sigma_{\rm{ST}}$ at early times may reflect more rapid and stochastic gas accretion in the early Universe.

\subsection{Investigating the long-term variability of the SFMS}
\begin{figure*}
    \centering	\includegraphics[width=2\columnwidth]{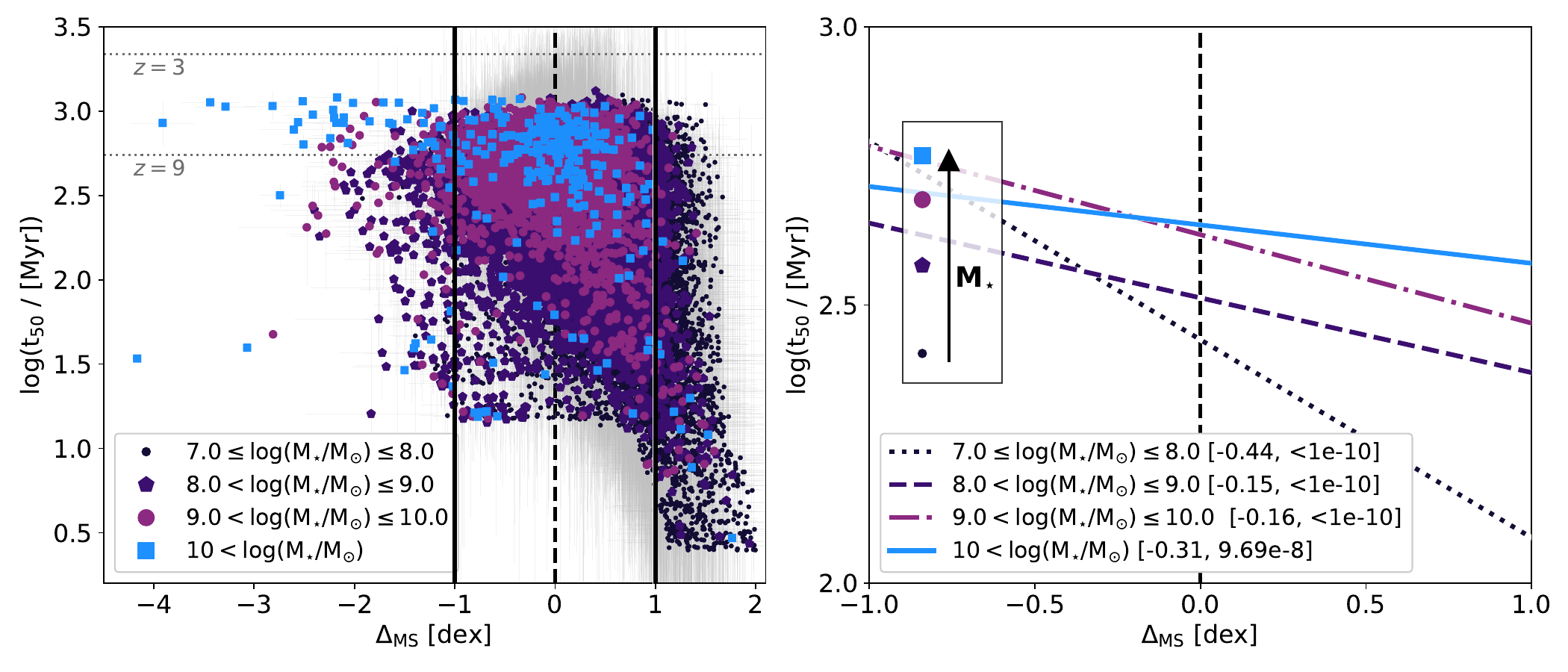}
    \caption{Relation between stellar age (e.g., half-mass assembly time, t$_{50}$) and distance from the SFMS, $\Delta_{\rm{MS}}$, using sSFR$_{10}$. \textsl{Left panel:} scatter plots for different stellar mass bins, as indicated in the legend. The lower cut-off in t$_{50}$ is due to the minimum look-back time in the SFH (5 Myr). The vertical dashed line shows $\Delta_{\rm{MS}}=0$, while the vertical filled lines delimit the region where $-1\leq\Delta_{\rm{MS}}\leq 1$. The horizontal dotted lines show the age of the Universe at $z=3$ and $z=9$. There are two tentative trends that arise in this figure. Firstly, a relation between the position of the galaxies on the SFMS and their age, suggesting that (on average) younger galaxies lie above the SFMS. Secondly, as stellar mass increases, galaxies are typically older and can reach lower $\Delta_{\rm{MS}}$ values (possibly due to quenching). \textsl{Right panel:} best fit lines to each stellar mass bin considered, for galaxies within the  vertical filled black lines in the left panel, to illustrate the trend between age, stellar mass, and $\Delta_{\rm{MS}}$. For each stellar mass bin, we provide in brackets the Spearman's rank coefficients in the format [correlation, p-value]. For reference, the correlation coefficient can adopt values between -1 and 1, where 0 represents no association between parameters, and a p-value close to zero indicates that the correlation -- even if weak-- is unlikely to be due to chance.  We note that the steepness of the slope for galaxies with log(M$_{\star}$/M$_{\odot}$)$=7.0-8.0$ is potentially due to stellar mass incompleteness in that bin. Finally, the markers show the median t$_{50}$ value per mass bin, confirming that, on average, galaxies become more massive as they age. Importantly, the overall correlation between stellar age and SFMS offset is weak, as indicated by the Spearman's rank coefficients, suggesting that short-term fluctuations in star formation dominate over longer-term evolutionary trends.}  
    \label{fig:t50_deltaMS}
\end{figure*}

In the previous analysis of the SFMS scatter, we have assumed that individual galaxies evolve about the same SFMS and the scatter is mainly caused by fluctuations on timescales of 10s of Myr. We now look into whether galaxies evolve about the SFMS on even longer timescales (e.g., 100s of Myr), which basically would imply that galaxies do not evolve around a single SFMS \citep{Patel2009,Gladders2013,Kelson2014,Abramson2016,Paulino-Afonso2020,Wang2020}. In order to assess this, we follow \citet{Abramson2016}, \citet{Wan2024} and \citet{Wan2025} and plot the inferred stellar age, as measured by the half-mass assembly time, $t_{50}$, as a function of the distance of the SFMS, $\Delta_{\rm MS}$, in Figure~\ref{fig:t50_deltaMS}.

The left panel of Figure~\ref{fig:t50_deltaMS} shows the relation between $t_{50}$ and $\Delta_{\rm MS}$ for individual galaxies in our sample, with colours indicating different stellar mass bins. Before interpreting these trends, we caution that $t_{50}$ is a particularly challenging quantity to constrain. This is especially true for galaxies with rising SFHs, where the light from younger stellar populations can outshine the contribution of older stars \citep{Papovich2023, Tacchella2023, Endsley2024, Narayanan2024, Witten2025}, leading to large uncertainties in the inferred stellar ages. 

We identify two tentative trends in the $t_{50}-\Delta_{\rm MS}$ plane. First, $t_{50}$ appears to be anticorrelated with $\Delta_{\rm MS}$, such that, on average, galaxies lying above the SFMS tend to be younger. Similar behaviour has been reported in previous studies \citep[e.g.][]{Langeroodi2024, Rinaldi2025, Wan2025}. However, this correlation is very weak, as indicated by the best-fit relation shown in the right panel of Figure~\ref{fig:t50_deltaMS}. For galaxies with log(M$_{\star}$/M$_{\odot}$)$> 8.0$, the median stellar age changes by less than a factor of two when moving from 1 dex below to 1 dex above the SFMS. The trend becomes significant only for the lowest-mass galaxies, where the age differences are more pronounced, but note that the steepness of the lower-mass bin is potentially due to stellar mass incompleteness. 

Secondly, when dividing our sample into stellar mass bins, we find that more massive galaxies tend to have older stellar populations. The median ages of each mass bin, shown in Figure~\ref{fig:t50_deltaMS}, further support the conclusion that higher-mass galaxies are typically older. In addition, the fraction of galaxies with low \(\Delta_{\rm MS}\) increases with stellar mass, consistent with a larger proportion of massive systems undergoing quenching. 

In summary, we find a broadly consistent picture in which younger galaxies tend, on average, to lie above the SFMS, likely reflecting the impact of recent starburst episodes in these systems. However, the overall correlation between stellar age and SFMS offset is weak, suggesting that short-term fluctuations in star formation dominate over longer-term evolutionary trends. 

\begin{figure*}
    \centering	\includegraphics[width=2\columnwidth]{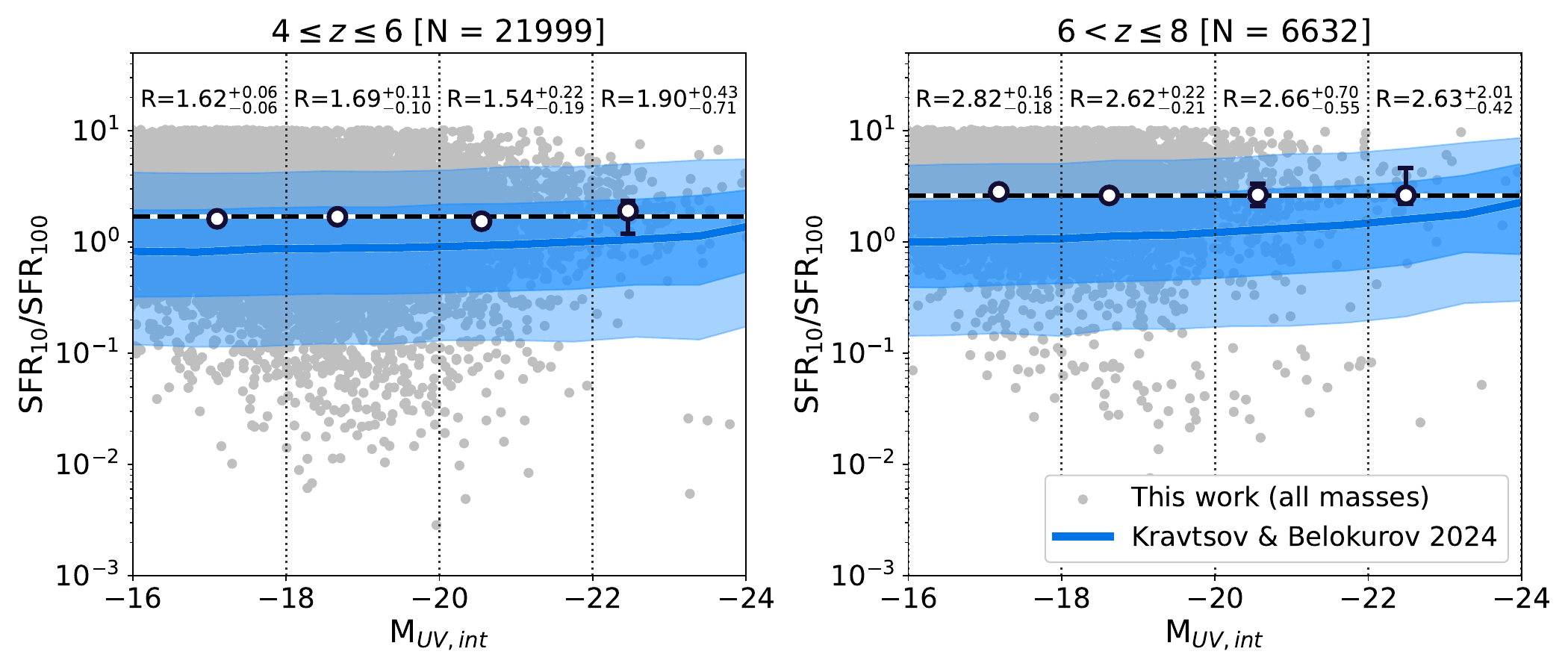}
    \caption{Ratio of star formation rates averaged over 10 and 100 Myr, respectively (R$\equiv$SFR$_{10}$/SFR$_{100}$), as a function of intrinsic UV magnitude. The scatter plots show the median measurements for galaxies in our sample, the upper limit of R=10 is given by the maximum SFR per each timescale ($\frac{1}{\text{t}_{\rm{average}}}$).
    The white circles with black edges show the median values for each UV magnitude bin (delimited by vertical dotted lines), the error bars reflect the number of galaxies in each bin and are obtained via bootstrapping. The median SFR$_{10}$/SFR$_{100}$ values increase as a function of redshift, as highlighted by the horizontal dashed lines (and corresponding values of the ratio). The blue filled curves and shaded areas represent the models from \citet{Kravtsov2024}, with their 68 and 98 percentiles , computed at $z=5$ and $z=7$, respectively. We find that the models are able to reproduce our \texttt{Prospector}-inferred values.}
    \label{fig:Kravtsov_burstiness}
\end{figure*}

\begin{figure}
    \centering	\includegraphics[width=1\columnwidth]{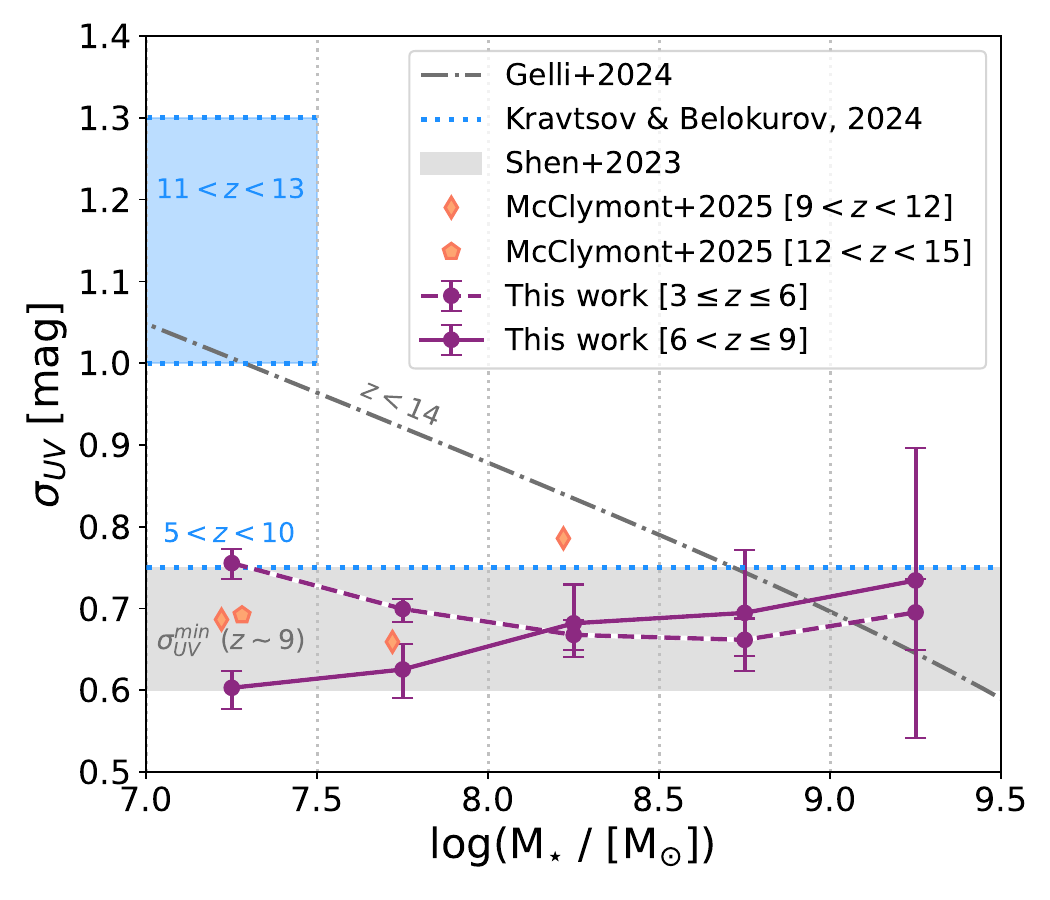}
    \caption{UV variability as a function of stellar mass for two redshift bins: $3\leq z\leq6$ (purple dashed) and $6<z\leq9$ (purple filled). Our measurements have been corrected for dust attenuation and the SED errors have been subtracted in quadrature. The UV traces longer timescales, and as such, is less sensitive to the stellar mass limits used. The grey shaded area shows the minimum UV variability needed to reproduce observations at $z\approx9$ from \citet{Shen2023}, its lower (upper)  bound represents dust corrected (uncorrected) values. The blue dotted horizontal lines and region show the $\sigma_{\rm{UV}}$ needed to reproduce UV luminosity functions at $5<z<10$ and $11<z<13$ from \citet{Kravtsov2024}, when a moderate amount of stochasticity in star formation is introduced. The  diamonds and pentagon show the $\sigma_{\rm{UV}}$ values for $9<z<15$ from \citet{McClymont2025}. Finally, the dot-dashed grey line shows the analytic scatter estimated by \citet{Gelli2024}, that was shown to reproduce the JWST observations up to $z=14$, when \textsc{THESAN-ZOOM} star formation efficiencies are adopted \citep{Shen2025}. 
    We find that our dust corrected UV variability converges at $\approx0.75$, in tight agreement with the $\sigma_{\rm{UV}}^{\rm{min}}$ from \citet{Shen2023} and the models from \citet{Kravtsov2024}.  However, more UV variability is required at lower stellar masses in order to explain the observed over-abundance of UV-bright galaxies at $z>10$.}
    \label{fig:scatter_UV}
\end{figure}

\section{Consequences of bursty star formation across cosmic time}
\label{SEC:implications}
We have shown that the SFMS can be parametrised up to $z=9$, and that SFHs are in general bursty across cosmic time. In this section we explore the consequences and implications of this result on three key points: (1) the observed over-abundance of UV-bright galaxies at $z>10$, (2) the considerations that should be taken into account when estimating the mass completeness limits of samples, and (3) the connection between bursty star formation and the reionisation of the Universe. 

\subsection{Implications at z>10}
\label{sec:implications_UVLF}
Since the launch of the JWST, several galaxies have been observed at $z>10$ \citep[][]{Naidu2022,Castellano2022,Curtis-Lake2023,Donnan2023,Finkelstein2023,Harikane2023,Robertson2023,Casey2024,D'Eugenio2024,Hainline2024,Harikane2024,McLeod2024,Wang2024,Helton2025,Kokorev2025,Naidu2025}. These observations have challenged our understanding of galaxy evolution, since measurements of the UV luminosity function \citep[e.g., ][]{Finkelstein2023,Adams2024,Donnan2024,Robertson2024, Whitler2025} have revealed a potential over-abundance of UV bright galaxies in the early Universe, when compared to model predictions. Some ways to reconcile theory with observations are to invoke significantly increased burstiness in galaxies at $z>9$, high star formation efficiencies and suppressed feedback in these galaxies \citep{Dekel2023,Li2024}, low dust attenuation as redshift increases \citep{Ferrara2023}, AGN activity \citep[which has been detected at $z>10$, see e.g, ][]{Maiolino2024,Scholtz2024}, or a top-heavy IMF \citep{Trinca2024,Mauerhofer2025}. Our focus here is variations in SFRs that arise from bursty SFHs can lead to strong UV variability which could, in turn, explain the observed over-abundance of UV-bright galaxies at $z>10$. In brief, high-redshift galaxies could be up-scattered due to a recent burst in star formation, making them bright enough to dominate the bright-end of the UV luminosity function \citep{Ren2019,Mason2023,Mirocha2023,Munoz2023,Shen2023}. 

At $z<9$, galaxies have SFHs that are \textit{consistent} with being bursty \citep[e.g., ][]{Looser2023,Endsley2023,Asada2024,Simmonds2024ELGs,Baker2025,Cole2025}. In this work, we explicitly demonstrate that galaxies at $z = 3-9$ exhibit bursty SFHs, as evidenced by excess star formation variability on 10 Myr timescales compared to 50 Myr. Furthermore, we provide a quantitative measurement of this variability, as shown in Figure~\ref{fig:scatter_short_timescales}. In order to assess whether our inferred star-formation variability is enough to explain the abundance of UV bright galaxies at $z>10$, we perform a comparison with the model by \cite{Kravtsov2024}, where stochasticity of star formation is introduced in order to reproduce the observed UV luminosity functions at $z\approx11-13$ and $z\approx 16$. They find that their model can successfully reproduce stellar masses and UV luminosities at those redshifts, after adopting only a modest lever of stochasticity of SFRs. Importantly, their models predict that the ratio between the SFRs averaged over 10 and 100 Myr (R$\equiv$SFR$_{10}$/SFR$_{100}$) should increase with redshift. This increase can be interpreted as galaxies becoming more bursty in the early Universe, and having on average rising SFHs. Figure~\ref{fig:Kravtsov_burstiness} shows the predictions from \cite{Kravtsov2024}, computed at $z=5$ and $z=7$, along with galaxies in redshift bins around these values, as indicated in the title of each panel. The grey circles show our \texttt{Prospector}-inferred measurements for all the galaxies in our original sample (i.e. without making a cut in stellar mass). The white points and error bars show the medians per UV magnitude bin (from M$_{\rm{UV}}=-16$ to $-24$), plotted at the median M$_{\rm{UV}}$ of each bin. The errors reflect the number of galaxies contained in each bin. Finally, the blue curve and the shaded areas represent the extrapolated models with their corresponding 68 and 95 percentiles. As can be seen in the figure, our median \texttt{Prospector}-inferred measurements per M$_{\rm{UV}}$ bin are within the range of the model predictions. Moreover, the models predict an increase in R from  $z=5$ to $z=7$ by a factor of $\approx1.2\pm0.2$, while our measurements of R increase by a factor of $\approx1.6\pm0.5$. Therefore, we find that the predictions from \cite{Kravtsov2024} are consistent with this work at $z<9$.  

We investigate the UV variability by measuring $\sigma_{\rm{UV}}$, which describes the scatter in M$_{\rm{UV}}$ at fixed stellar mass. It should be emphasised that $\sigma_{\rm{UV}}$ can encapsulate information beyond burstiness of star formation. For example, it includes the scatter in the M$_{\star}$-M$_{\rm halo}$ conversion (since in some models $\sigma_{\rm{UV}}$ is defined at fixed halo mass), the long-timescale diversity of SFHs at a fixed stellar mass, and the translation from SFHs to UV luminosities. The latter contains possible scatter from metallicity, nebular contribution, IMF variations, SFH assumptions, and, importantly, dust attenuation. Therefore, a large $\sigma_{\rm{UV}}$ does not necessarily mean burstier SFHs, since it can arise from several other factors.

Figure~\ref{fig:scatter_UV} shows the dust-corrected $\sigma_{\rm{UV}}$ as a function of stellar mass for two redshift bins, $3\leq z\leq6$ and $6<z\leq 9$.
The grey shaded area represents the minimum $\sigma_{\rm{UV}}$ from \cite{Shen2023}, needed to match observational constraints at $z\sim9$, and the lower and upper boundaries show $\sigma_{\rm{UV}}$ with and without dust corrections, respectively. For comparison, and motivated by the agreement of the \cite{Kravtsov2024} models with our galaxy properties, we include the $\sigma_{\rm{UV}}$ needed in their work in order to reproduce the observed UV luminosity functions at $5<z<10$ ($\sigma_{\rm{UV}}\approx0.75$), and at $11<z<13$ ($\sigma_{\rm{UV}}\approx1-1.3$), as blue dotted lines and shaded area. We find $\sigma_{\rm{UV}}\approx 0.75$ for our dust-corrected measurements in both redshift bins, with little-to-no dependence on stellar mass. This result agrees remarkably well with the minimum $\sigma_{\rm{UV}}$ from \cite{Shen2023} and the predictions including stochastic star formation from \cite{Kravtsov2024}, as well as the simulation-based studies from \cite{Pallottini2023} and \cite{Semenov2024}, in our redshift range \citep[for a compilation of simulation results for high-redshift galaxies as a function of halo mass, see Figure 4 of ][]{Shen2024}. Moreover, \cite{McClymont2025} use \textsc{THESAN-ZOOM} simulated galaxies to measure the UV variability up to $z=15$ (without accounting for dust attenuation), and find that at $z>6$ their $\sigma_{\rm{UV}}$ converges to $\approx0.75$. Importantly, if we combine our estimations at $3<z<9$ with their measurements at $9<z<15$, we find a consistent picture of $\sigma_{\rm{UV}}\approx 0.75$, with virtually no redshift evolution \citep[a result that is also in agreement with other studies, e.g., ][]{Ciesla2024,Shuntov2025}. \cite{McClymont2025} explain this behaviour due to the UV tracing long-term scatter, which is only introduced as cosmic time increases.

To understand if the JWST observations at $z>10$ can be explained by the UV variability, we show the analytic scatter estimated by \cite{Gelli2024}  (grey dot-dashed line), which is needed to reproduce UV luminosity functions up to $z=14$ when adopting the \textsc{THESAN-ZOOM} star formation efficiency, as shown in \cite{Shen2025} . 
A comparison between our estimated $\sigma_{\rm{UV}}$ and the measurements from \cite{McClymont2025}, with the \cite{Shen2025} limit, suggests that additional scatter is needed at lower stellar masses to explain the over-abundance of observed UV-bright galaxies at $z>10$. Thus, we find that the UV variability we measure at $z=3-9$ is not enough to explain the UV luminosity functions at $z>10$ \citep[see also][]{Carvajal-Bohorquez2025}. Possible ways to address this issue include, for example, to invoke a top-heavy IMF \citep{Shen2023,Mauerhofer2025}, variable star formation efficiencies \citep{Dekel2023,Somerville2025}, or a UV variability that evolves strongly with redshift \citep{Kravtsov2024}. Additionally, our measurements at the lowest stellar masses may underestimate the true effect due to sample incompleteness. 

\begin{figure*}
    \centering	\includegraphics[width=2\columnwidth]{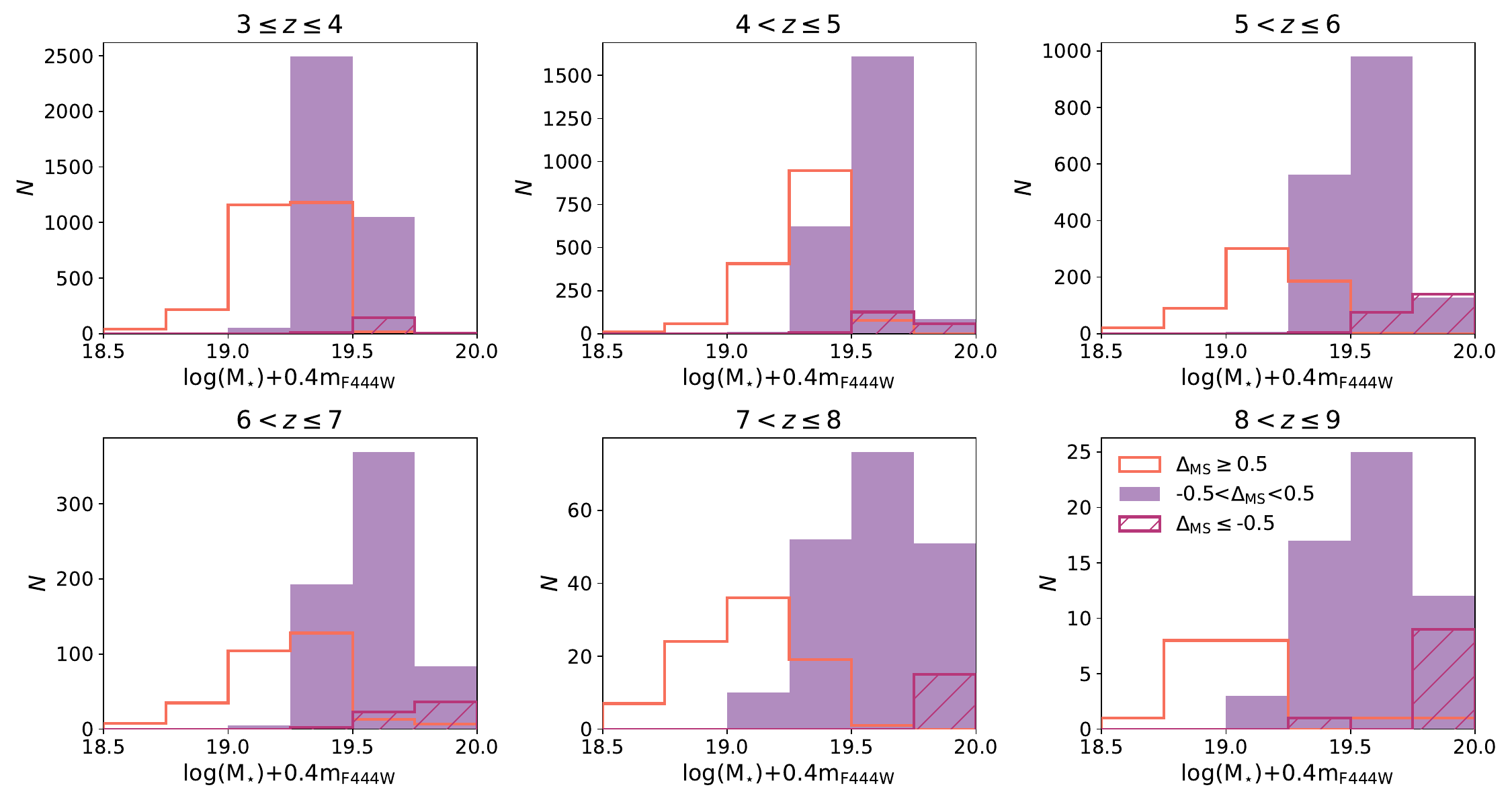}
    \caption{Mass-to-light ratio for galaxies with log(M$_{\star}$/M$_{\odot}$)$=8.0-8.5$, per redshift bin. In each panel we divide galaxies into three bins depending on their location respective to the SFMS: above ($\Delta_{\rm{MS}}\geq0.5$, orange), on ($-0.5 < \Delta_{\rm{MS}}<0.5$, purple filled), and below ($\Delta_{\rm{MS}}\leq-0.5$, red hatched) the relation, respectively. We find a clear dependence of the mass-to-light ratio on $\Delta_{\rm{MS}}$, where galaxies above the SFMS are brighter. Therefore, the assumption that the galaxies at the magnitude limit have the same mass-to-light ratio as the galaxies below it might be too simplistic, explaining why the region with log(M$_{\star}$/M$_{\odot}$)$\approx8.1-9.0$ is only partially complete.}
    \label{fig:M2L}
\end{figure*}

\subsection{Implications on stellar mass completeness}
\label{sec:implications_mass}

In Section~\ref{sec:completeness}, we estimate the stellar mass completeness of our sample assuming constant mass-to-light ratios for a fixed F444W magnitude. We argue that this is a lower limit, and define partially complete and complete regions based on the shape of the sSFR as a function of stellar mass (see Figure~\ref{fig:mass_completeness_true}). By analysing the scatter of the SFMS, we find that SFHs are bursty on average, but more so at low stellar masses. Therefore, in this section, we explore the implications that bursty star formation has on estimations of stellar mass completeness.   

We have shown in Figure~\ref{fig:flux_comparisons} that \texttt{Prospector} can predict \ha\/ and \oiii\/$_{\lambda5007}$ emission line fluxes for a subsample of galaxies with JWST NIRSpec measurements from JADES, which means that recent SFRs can also be reliably estimated for these galaxies. Although promising, this result is based on a subsample composed of galaxies with detectable emission lines, and as such, is not representative of the entire galaxy population. Before making conclusions about stellar mass completeness methods, we first investigate if the partially complete region is due to \texttt{Prospector} limitations. Specifically, we consider if the code struggles with retrieving SFHs (and therefore, SFRs) when stellar masses and SFRs are low. In order to test this, we produce mock observations for two galaxies with log(M$_{\star}$/M$_{\odot}$)$\approx7.5$, with no recent star formation, and at the extremes of our redshift range. We decide to select a stellar mass below our derived stellar mass completeness limits to ensure the reliability of our method. The results are shown in Appendix~\ref{app:SFR_retrieval}, and demonstrate that \texttt{Prospector} can indeed recover declining SFHs for low-mass galaxies even in the absence of medium band photometry. Therefore, the partially complete region cannot be explained by limitations in our SED modelling routine.

After demonstrating that \texttt{Prospector} can retrieve galaxy properties remarkably well at the masses and redshifts used in this work, we now focus on the behaviour of the mass-to-light ratio of the galaxies in our sample, for a given stellar mass. Figure~\ref{fig:M2L} shows this ratio for galaxies with log(M$_{\star}$/M$_{\odot}$)$=8.0-8.5$, which is well represented at all redshifts studied here. For each redshift, we divide galaxies into three bins depending on their distance to the SFMS. In particular, we show galaxies above ($\Delta_{\rm{MS}}\geq0.5$), on ($-0.5 < \Delta_{\rm{MS}}<0.5$), and below ($\Delta_{\rm{MS}}\leq-0.5$) this relation. We find that the mass-to-light ratio varies as a function of $\Delta_{\rm{MS}}$ for a given stellar mass, and thus, that the assumption that the galaxies at the magnitude limit have the same mass-to-light ratio as the galaxies below it might be too simplistic (Section~\ref{sec:completeness}). As a consequence, the stellar mass completeness limits derived under this assumption are optimistic and should be used with caution, especially when star formation is bursty in nature \citep{Sun2023}.  

The process of fitting the SFMS is highly sensitive to the stellar mass completeness of the sample. If low-mass galaxies with low SFRs are missed, then the overall effect is that sSFRs depend negatively on stellar mass, since the relation becomes dominated by low-mass galaxies with recent star formation. This is illustrated by the piece-wise fits in Figure~\ref{fig:MS_fit_10_complete}, which show that the slope of the relation between sSFR and stellar mass becomes negative when adopting the stellar mass limits derived following \cite{Pozzetti2010}. \cite{McClymont2025} study the implications that this observational bias has on the shape and evolution of the SFMS. Interestingly, they find that when the stellar mass completeness is not properly assessed, the SFMS evolution with redshift is less steep, mimicking the flattening seen in $z<6$ observational studies such as \cite{Speagle2014} and \cite{Popesso2023}. 

In this work, we fit the SFMS in the stellar mass range where the sSFR for each timescale is flat (see Figure~\ref{fig:mass_completeness_true}). Presenting an alternative mass completeness estimation method is outside of the scope of this study, however, we caution future observational studies to look carefully into this subject when fitting the SFMS. Thanks to facilities such as the JWST, and large surveys like JADES, having statistical samples of galaxies in the early Universe is now possible. 

\subsection{Implications for reionisation} 
\label{sec:implications_EoR}
One of the major ongoing debates in cosmology is pinpointing which sources are mainly responsible for ionising the Universe by $z\sim5-6$: star-forming galaxies or AGN. Before the launch of the JWST, the community widely agreed on young stars in star-forming galaxies as being the dominant source of ionising photons in the early Universe \citep[e.g., ][]{Hassan2018,Rosdahl2018,Trebitsch2020,Chisholm2022}. However, ever since the launch of this telescope, a population of low luminosity AGN has been unveiled \citep[][]{Juodzbalis2023,Maiolino2024,Ubler2024}, reigniting this debate. Although most works find that AGN are subdominant to the reionisation process \citep[][]{Asthana2024,Dayal2024,Jiang2025}, some find the opposite \citep[see for example, ][]{Grazian2024,Madau2024}. In the context of star-forming galaxies as the main agents of reionisation, faint low-mass galaxies with bursty SFHs have been proposed to be key agents during the EoR \citep[e.g., ][]{Seeyave2023,Rinaldi2024,Simmonds2024ELGs,Simmonds2024complete,Choustikov2025}, but see also \cite{Naidu2020}. It is important to highlight that for galaxies to be able to ionise the Universe, the LyC radiation produced in their interiors must escape the interstellar medium to reach the IGM. High LyC escape fractions (f$_{\rm{esc}}$) have been directly observed \citep[e.g., ][]{Borthakur2014,Vanzella2018,Izotov2021,Kerutt2024}, but usually not in large samples \citep{Leitet2013,Leitherer2016,Steidel2018,Flury2022}. Moreover, this radiation can only be measured up to $z\sim5$ due to the increasingly neutral IGM \citep{Madau1995,Inoue2014}, making f$_{\rm{esc}}$ the most elusive ingredient in studies regarding the sources responsible for the EoR. 

A common way to investigate which galaxies are responsible for the reionisation of the Universe is through the ionising photon production efficiency, \xion\/, given by the ratio between the ionising photon flux and the non-ionising UV continuum luminosity. Given the challenges in measuring ionising photons in galaxies, \xion\/ is usually estimated by assuming Case-B recombination (i.e. no escape of ionising radiation, \xionnofesc\/). This property has been thoroughly investigated in the literature \citep[e.g., ][]{Bouwens2016,Faisst2019,Endsley2023,Stefanon2022,Prieto-Lyon2023}. Not surprisingly, \xionnofesc\/ correlates with SFR$_{10}$/SFR$_{100}$ \citep{Simmonds2023}. In Figure~\ref{fig:MS_fit_10_complete}, we show the best fit to the SFMS using the SFR averaged over the past 10 Myr, and colour-coded by \xionnofesc\/. The resulting colour gradient that can be seen indicates that this property also correlates with the distance to the SFMS, $\Delta_{\rm{MS}}$, where the galaxies that are more efficient producers of ionising radiation are also the ones that lie above the SFMS, and are likely the type of galaxy dominating the ionisation of the Universe. 

Tying this work to the results from \cite{Simmonds2023} and \cite{Simmonds2024complete}, through our study of the SFMS, we confirm that galaxies with stellar masses log(M$_{\star}$/M$_{\odot}$)$=8.0-10.0$ have bursty star formation. Moreover, we find that galaxies with $8.0\leq$log(M$_{\star}$/M$_{\odot}$)$\leq8.5$ have a higher short-term scatter, indicating increased burstiness. Therefore, our work supports the scenario of low-mass galaxies with bursty star formation being responsible for the bulk of the ionising photon budget in the early Universe. We expect our results can be extrapolated to lower stellar masses, however, we caution that we are not complete below log(M$_{\star}$/M$_{\odot}$)$\sim8.0$. Finally, in light of our new insights into stellar mass completeness, we re-fit the \xion\/ relations from \cite{Simmonds2024complete} using only galaxies with log(M$_{\star}$/M$_{\odot}$)$\geq9.0$ and find that the main conclusions of the work remain unchanged within errors, where \xion\/ depends on redshift and M$_{\rm{UV}}$ as:
\begin{multline*}
\log(\xi_{\rm{ion}} (z,\text{M}_{\rm{UV}})) = \\ 
    (-0.003\pm 0.015)z + (-0.019\pm 0.015)\text{M}_{\rm{UV}} + (24.882\pm 0.310).
\end{multline*}
This indicates that even though the SFMS is extremely sensitive to the stellar mass completeness of the sample, studies on the ionising properties of galaxies can be performed using less conservative stellar mass completeness estimates.

\section{Caveats and limitations}
\label{SEC:caveats}
The first and most important source of uncertainty in this work comes from the use of SED modelling, and the properties that are derived from this process. Importantly, it is challenging to measure individual SFHs, since to do so it is necessary to break degeneracies between indicators of burstiness (such as SFR$_{10}$/SFR$_{100}$) from parameters like dust attenuation, metallicity, and ionisation parameter \citep{Johnson2013,Broussard2019}. Although, we note that mid-infrared observations obtained with MIRI can aid in estimating physical properties of galaxies at high-redshift star-forming galaxies \citep{Helton2025SED}.  In a recent work, \cite{Wang2025} study the recovery of SFRs when adopting different SFH priors, and find that although SFRs averaged over the past 30 Myr can be reasonable well recovered (with a median offset of $\sim0.1$ dex), that fluctuations at shorter timescales are not well captured, and that bursty star formation must be taken into account. Besides these complications, SED fitting codes struggle to find reliable SFHs at $z>7$ when older stellar populations are outshone by recent star formation  \citep{Papovich2023,Tacchella2023,Endsley2024,Narayanan2024,Witten2025}. 

In addition to these difficulties, a prior for the SFH must be assumed. We adopt a non-parametric continuity prior \citep{Leja2019}, consisting of 8 bins of star formation, where the ratio between these bins is fit with a Student's t-distribution prior with a scale of 0.3. The latter means that some burstiness is allowed between adjacent bins, and that both star-forming and quenched galaxies can be reproduced \citep{Leja2019,Johnson2021,Haskell2023}, but that extreme solutions are not preferred. As a result, the burstiness found in this work might be underestimated, although the exact implications of using this SFH are not here explored. Promisingly, the continuity SFH used in this work has been shown to be less prone to outshining \citep{Wang2025}. Besides all the caveats introduced by the SED fitting process itself, as in \cite{Simmonds2024complete}, we assume a Chabrier IMF, with a maximum stellar mass of 100 M$_{\odot}$. This choice of IMF was motivated by the size of the sample, in the hope of selecting a prescription that represents the whole dataset. However, this IMF might not be the best option if more extreme stellar populations are required \citep[for example, a top heavy IMF, as in ][]{Cameron2024}, if the IMF depends on the physical properties of the gas in each individual galaxy \citep{Haslbauer2024,Kroupa2024,Zonoozi2025}, or if the IMF evolves as a function of cosmic time.    

Secondly, as mentioned in the previous section, ever since the launch of the JWST, a significant population of low-luminosity AGN at high redshifts has been discovered \citep{Juodzbalis2023,Madau2024,Maiolino2024,Ubler2024,Scholtz2025}. We removed the JWST MIRI selected AGN from \cite{Lyu2024} and the broad-line AGN from \cite{Juodzbalis2025} to avoid contamination from these sources, however, it is possible that our sample contains other AGN, which would lead to our stellar masses being overestimated in these cases \citep{Buchner2024}. Unfortunately, it is virtually impossible to reliably identify AGN with photometry alone \citep{Wasleske2024}. As such, we caution the reader that our sample might contain AGN when their emission can be mimicked by stellar emission. 

Finally, our work relies on photometric redshifts. In particular, we use \texttt{EAzY} redshifts as priors for our \texttt{Prospector} fitting routine. The former has been proven to yield accurate results for large JADES samples using photometry as input \citep{Rieke2023}. In addition, the \texttt{Prospector}-inferred redshifts are in good (but not perfect) agreement with spectroscopic redshifts from literature \citep{Simmonds2024complete}. This indicates that overall our photometric redshifts are reliable, but that some outliers are inevitably part of this work. We stress that this outlier fraction is small (of the order of at most a few percent). Moreover, our imposed $\chi^2$ cut should ensure that the data is well reproduced by our best-fitting models. 

As a summary, although there are limitations to our methods since they rely on photometry, photometric redshifts, and SED modelling, they are -- as much as possible -- accurate and representative of the galaxy population at $3\leq z\leq9$.

\section{Summary and Conclusions}
\label{SEC:conclusions}

In this work, we have utilized NIRCam photometry of galaxies in the GOODS-N and GOODS-S fields at redshifts $3<z<9$, combined with galaxy properties inferred using the SED fitting code \texttt{Prospector}, to study the star-forming main sequence (SFMS) and its scatter. We define two regions based on stellar mass: a partially complete region (log(M$_{\star}$/M$_{\odot}$)$\approx 8.1-10.3$, with 18,344 galaxies) and a more conservative complete region (log(M$_{\star}$/M$_{\odot}$)$=9.0-10.3$, with 2989 galaxies). Given the sensitivity of the SFMS shape to the stellar mass completeness limit, we focus on fitting the SFMS in the complete region, extrapolating to lower masses when considering scatter.

Our main findings can be summarised briefly as:
\begin{itemize}
    \item The SED modelling code \texttt{Prospector} is able to retrieve emission line fluxes and SFHs reliably when adopting the continuity SFH prior, even in the absence of medium-band photometry.
    \item The redshift dependence of the SFMS at an averaging timescale of 10 Myr (sSFR$_{10}$) is $(1+z)^{\mu}$ with $\mu=2.30^{+0.03}_{-0.01}$, which is close to the specific mass accretion rate onto dark matter for a $\Lambda$CDM cosmology in the Einstein-deSitter regime \citep{Dekel2013} and overall consistent with predictions from numerical simulations \citep[e.g., ][]{Tacchella2016,McClymont2025}.
    \item The SFMS normalisation varies in a complex way with the SFR averaging timescale, reflecting the combined effects of bursty star formation and increased prevalence of rising SFHs toward higher redshifts. 
    \item We find that the sSFR is nearly independent of stellar mass (e.g., $\mathrm{SFR}\propto\mathrm{M}_{\star}^{\alpha}$ with $\alpha\approx1$) when completeness of the sample is taken into account, suggesting that galaxies with a variety of stellar masses double their mass on a similar timescale. Our SFMS agrees in shape and behaviour with the one inferred using the \textsc{THESAN-ZOOM} simulation \citep{McClymont2025}, suggesting that the flattening of the SFMS (e.g., $\alpha<1$) found in previous studies up to $z\approx 6$ is likely due to incompleteness in stellar mass. 
    \item Our analysis of the SFMS scatter indicates that galaxies in the mass range log(M$_{\star}$/M$_{\odot}$)$=8.0-10.0$ have bursty SFHs. Importantly, the short-term variability indicates that the burstiness is higher at lower stellar masses.
    \item We observe no significant trend of stellar age and stellar mass with distance to the SFMS. On average, we find that massive (low-mass) galaxies are slightly older (younger), and that older (younger) galaxies tend to lie below (above) the SFMS, consistent with short-term fluctuations in star formation dominating the scatter of the SFMS.
    \item We estimate the UV variability ($\sigma_{\rm{UV}}$) to analyse the impact of bursty SFHs on the observed over-abundance of UV-bright galaxies at $z>10$. We conclude that additional scatter must be introduced in order to explain these observations. This could be achieved by, for example, invoking a top-heavy IMF, variable star formation efficiencies, or significantly increasing the burstiness of star formation at $z>9$.
    \item We caution future studies about estimating stellar mass completeness limits that assume a constant mass-to-light ratio for a given stellar mass. We show that this ratio depends on the distance to the SFMS. Therefore, a conservative approach, such as the one adopted in this work, is suggested when studying the SFMS, 
\end{itemize}

Overall, we confirm in this work that SFHs are bursty across cosmic time, with lower-mass galaxies showing increased burstiness. In this work we aimed to answer two overarching questions: (1) How universal is the SFMS at early times? and, (2) What processes govern its emergence and evolution? We find that the SFMS is well in place up to $z=9$. Therefore, the emergence of the SFMS probably takes place at $z>9$. In order to push into this redshift regime, we would need a stellar-mass complete sample of galaxies at $z\approx9-14$. It is unclear if this will feasible in the near future, however, deeper imaging -- especially including MIRI coverage -- is needed \citep{Helton2025}.    

\section*{Data Availability}
The data underlying this article will be shared on reasonable request to the corresponding author. Fully reduced NIRCam images and NIRSpec spectra are publicly available on MAST (\url{https://archive.stsci.edu/hlsp/jades}), with \url{doi:10.17909/8tdj-8n28}, \url{doi:10.17909/z2gw-mk31}, and \url{doi:10.17909/fsc4-dt61} \citep{Rieke2023,Eisenstein2023JADES,Bunker2024,D'Eugenio2025}.

\section*{Acknowledgements}
CS thanks Ian Wolter and Gabriel Maheson for insightful discussions about star formation and dust. The JADES Collaboration thanks the Instrument Development Teams and the instrument teams at the European Space Agency and the Space Telescope Science Institute for the support that made this program possible. We also thank our program coordinators at STScI for their help in planning complicated parallel observations. This work is based [in part] on observations from JADES \citep{Rieke2023,Eisenstein2023JADES}, JEMS \citep{Williams2023}, and JOF \citep{Eisenstein2023JOF}, made with the NASA/ESA/CSA James Webb Space Telescope. The data were obtained from the Mikulski Archive for Space Telescopes at the Space Telescope Science Institute, which is operated by the Association of Universities for Research in Astronomy, Inc., under NASA contract NAS 5-03127 for JWST. These observations are associated with PIDs 1180, 1181, 120, 1286, 1895, 1963, and 3215. We would like to acknowledge the FRESCO team, led by Pascal Oesch, for developing their observing program with a zero-exclusive-access period. CS, FDE, and RM acknowledge support by the Science and Technology Facilities Council (STFC) and by the ERC through Advanced Grant number 695671 ‘QUENCH’, and by the UKRI Frontier Research grant RISEandFALL. RM also acknowledges funding from a research professorship from the Royal Society. WM and AS thank the Science and Technology Facilities Council (STFC) Center for Doctoral Training (CDT) in Data Intensive Science at the University of Cambridge (STFC grant number 2742968) for a PhD studentship. ECL acknowledges support of an STFC Webb Fellowship (ST/W001438/1). BDJ, JMH and ZJ acknowledge the JWST/NIRCam contract to the University of Arizona, NAS5-02015. AK was supported by the NASA ATP grant 80NSSC20K0512 and the National Science Foundation grant AST-2408267. BER acknowledges support from the NIRCam Science Team contract to the University of Arizona, NAS5-02015. WMB gratefully acknowledges support from DARK via the DARK fellowship. This work was supported by a research grant (VIL54489) from VILLUM FONDEN. VB acknowledges support from the Leverhulme Research Project Grant RPG-2021-205: ‘The Faint Universe Made Visible with Machine Learning’. AJB and JC acknowledge funding from the "FirstGalaxies" Advanced Grant from the European Research Council (ERC) under the European Union's Horizon 2020 research and innovation program (Grant agreement No. 789056). SC acknowledges support by European Union’s HE ERC Starting Grant No. 101040227 – WINGS. JMH is supported by JWST Program 3215. TJL gratefully acknowledges support from the Swiss National Science Foundation through a Postdoc.Mobility Fellowship and from the JWST Program 5997. The research of CCW is supported by NOIRLab, which is managed by the Association of Universities for Research in Astronomy (AURA) under a cooperative agreement with the National Science Foundation.



\bibliographystyle{mnras}
\bibliography{bib} 


\appendix
\section{Comparing star formation rates}
\label{app:SFR10_SFRHa}
\begin{figure}
    \centering	\includegraphics[width=1\columnwidth]{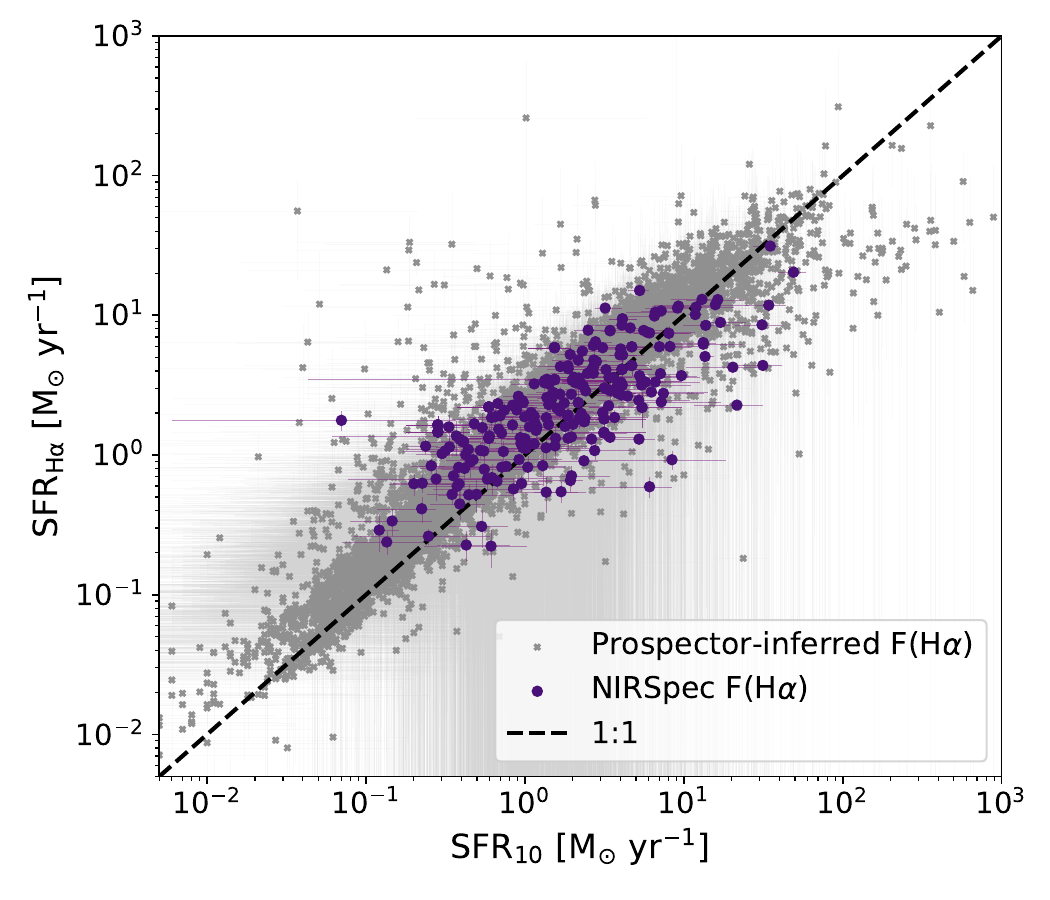}
    \caption{SFR$_{\rm{H}\alpha}$ versus SFR$_{10}$ for all galaxies in our sample with $8.0\leq \log(\text{M}_{\star}/\text{M}_{\odot})\leq10.3$ (grey, using \texttt{Prospector}-inferred H$\alpha$ fluxes), and for the subsample with NIRSpec measurements (purple). The dashed black line shows the 1:1 relation, and confirms that SFR$_{10}$ traces recent star formation, as seen by H$\alpha$ emission.}      \label{fig:app_SFR10_SFRHa}
\end{figure}

Comparison of the star formation rates averaged over the past 10 Myr (SFR$_{10}$) with those derived from the H$\alpha$ dust-corrected luminosities using the dust-corrected H$\alpha$ luminosity conversion from \cite{Shivaei2015SFR}. This SFR conversion was estimated based on the \cite{Salpeter1955} IMF, therefore, we apply a multiplicative factor of 0.64 to convert it to the \cite{Chabrier2003} IMF used in this work. Figure~\ref{fig:app_SFR10_SFRHa} shows this comparison using the \texttt{Prospector}-inferred H$\alpha$ fluxes for galaxies with $8.0\leq \log(\text{M}_{\star}/\text{M}_{\odot})\leq10.3$ (grey), and using the fluxes measured with NIRSpec  (purple, as seen in Figure~\ref{fig:flux_comparisons}). As can be see, there is an overall agreement between SFR$_{10}$ and SFR$_{\rm{H}\alpha}$.

\section{Mirroring distributions}
\label{app:mirror}
Process of mirroring non-Gaussian distributions of $\Delta_{\rm{MS}}$ for sSFR$_{10}$, described in Section~\ref{subsec:scatter_estimation}.
\begin{figure*}
    \centering	\includegraphics[width=2\columnwidth]{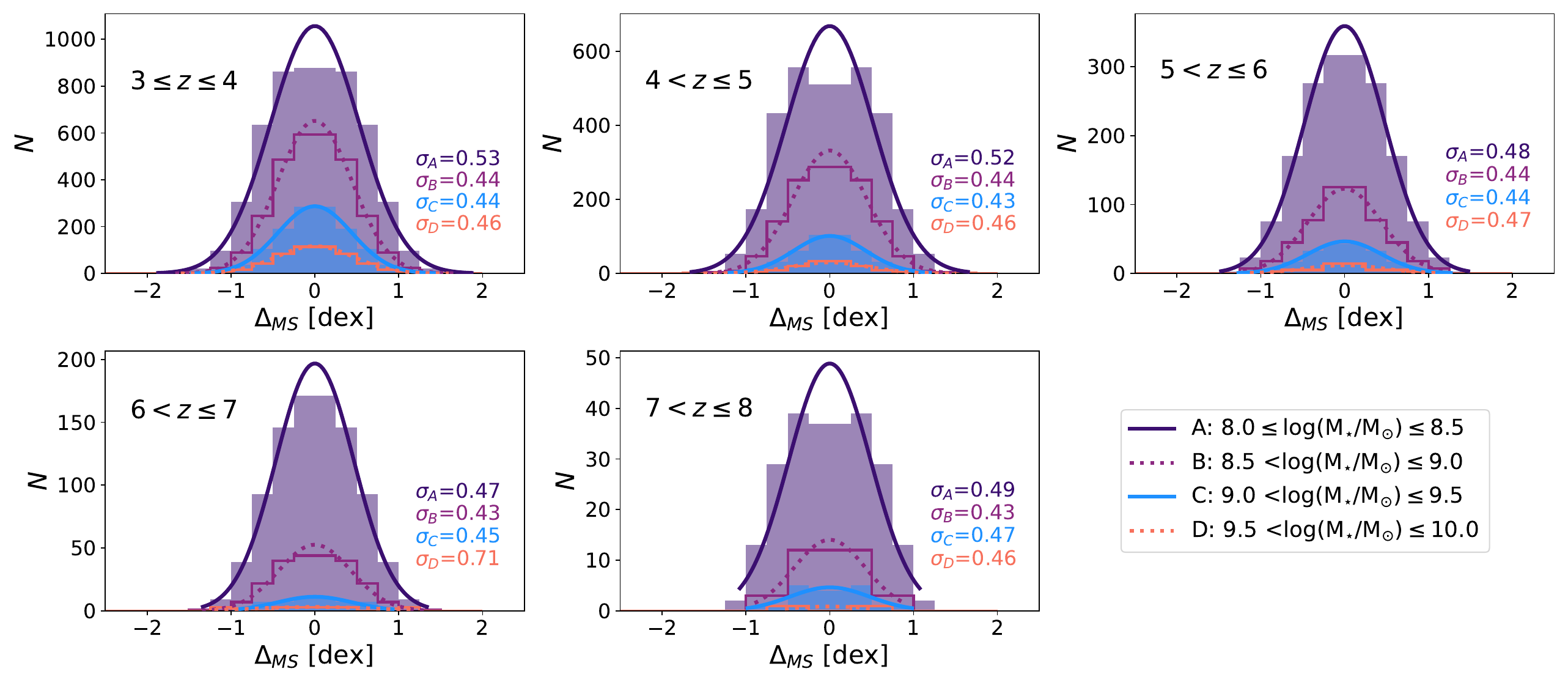}
    \caption{Same as in Figure~\ref{fig:scatter_10_linear} but mirrored around zero, in order for them to be described by Gaussian distributions. The  standard deviation for each mass bin is shown for each panel, labelled according to the legend. This scatter represents the median observed scatter, $\sigma_{\rm{obs}}$ (i.e., includes scatter introduced by the SED fitting routine). We note that for a given mass bin, $\sigma_{\rm{obs}}$ does not show a strong evolution with redshift.}      \label{fig:scatter_10_mirrored_linear}
\end{figure*}

\section{Star formation histories retrieval tests}
\label{app:SFR_retrieval}

\begin{figure*}
    \centering
    \includegraphics[width=1\linewidth]{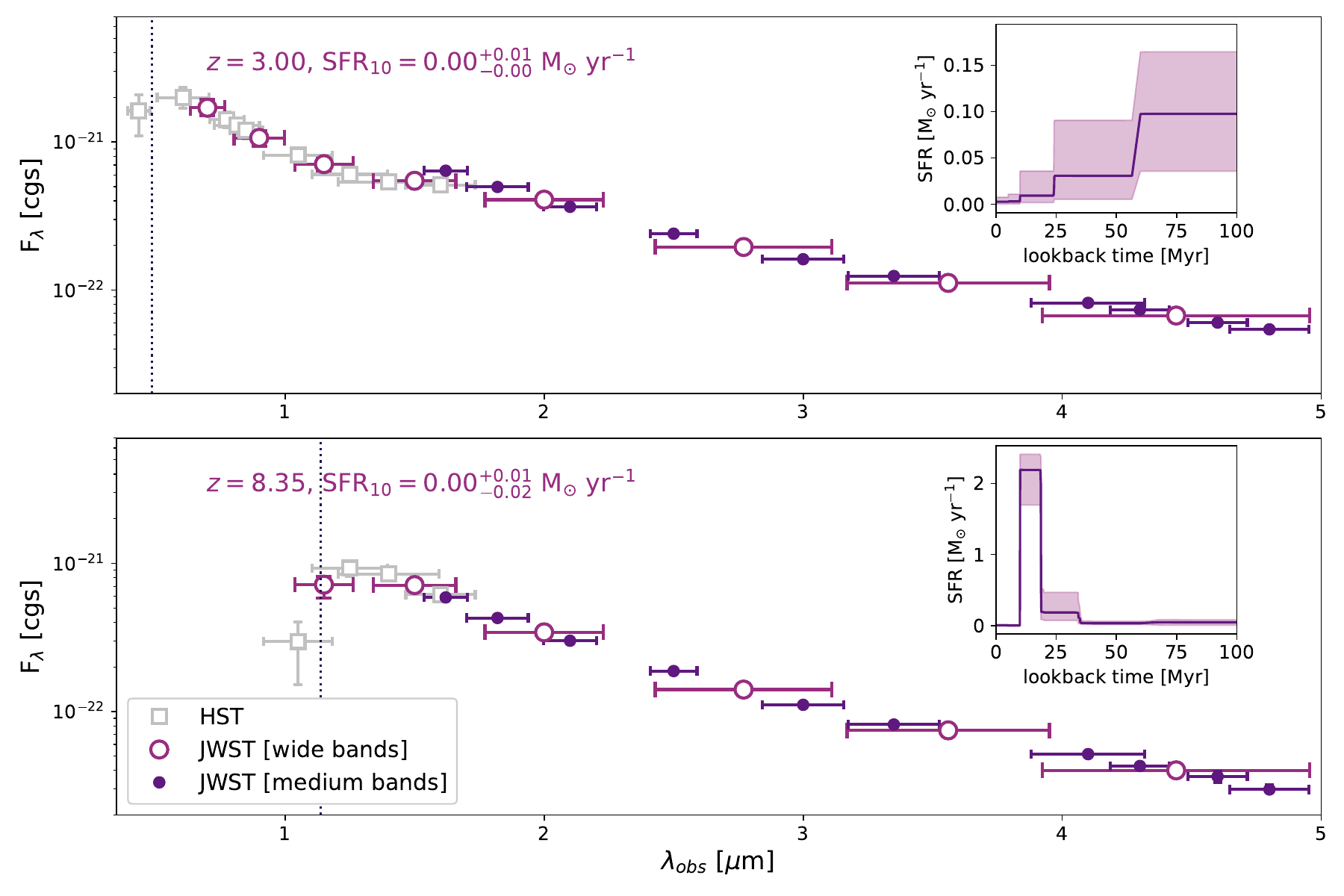}
    \caption{Input photometry and star formation histories for two galaxies with log(M$_{\star}$/M$_{\odot}$)$\approx7.5$, with no recent star formation, at $z=3.00$ (top) and $z=8.35$ (bottom), respectively. The different markers show photometry from HST (squares) and JWST wide (larger open circles) and medium (smaller filled circles) bands. We include the redshifted wavelength of \lya\/ for reference as a vertical dotted line.}
    \label{app:SFH_input}
\end{figure*}

\begin{figure*}
    \centering
    \includegraphics[width=0.3\linewidth]{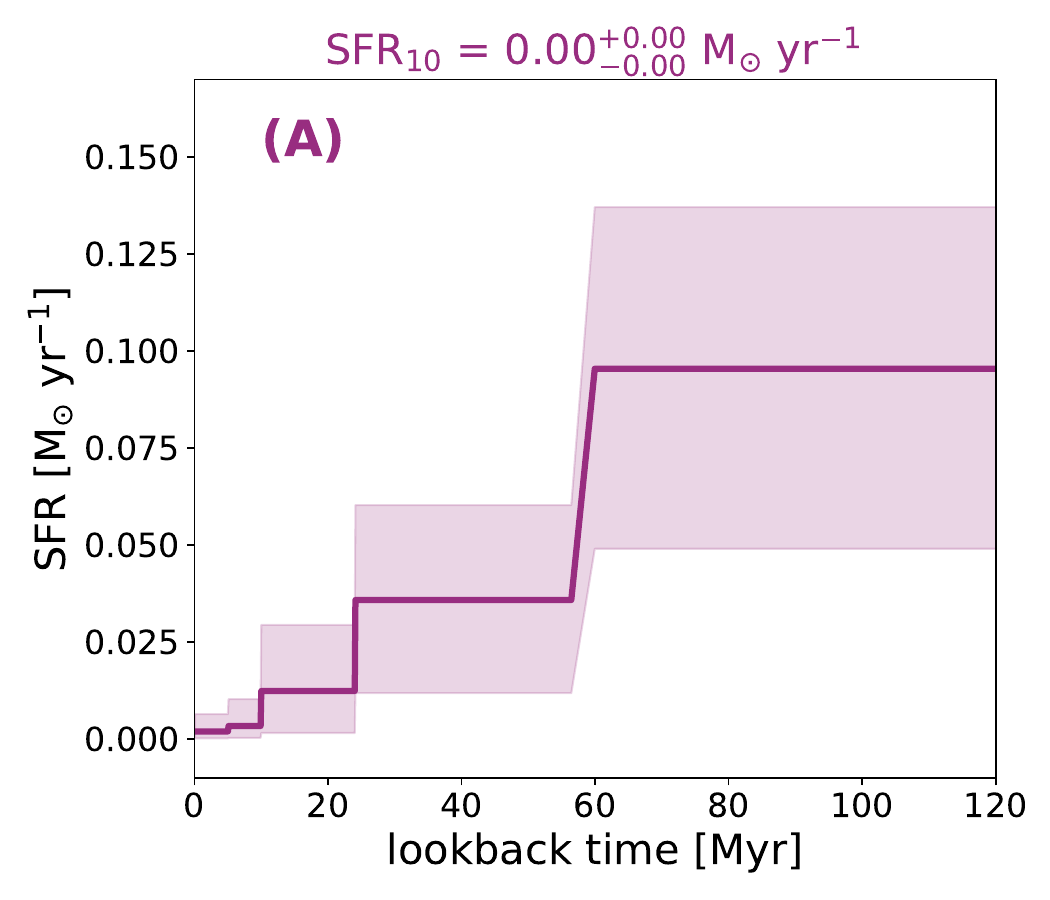}
    \includegraphics[width=0.3\linewidth]{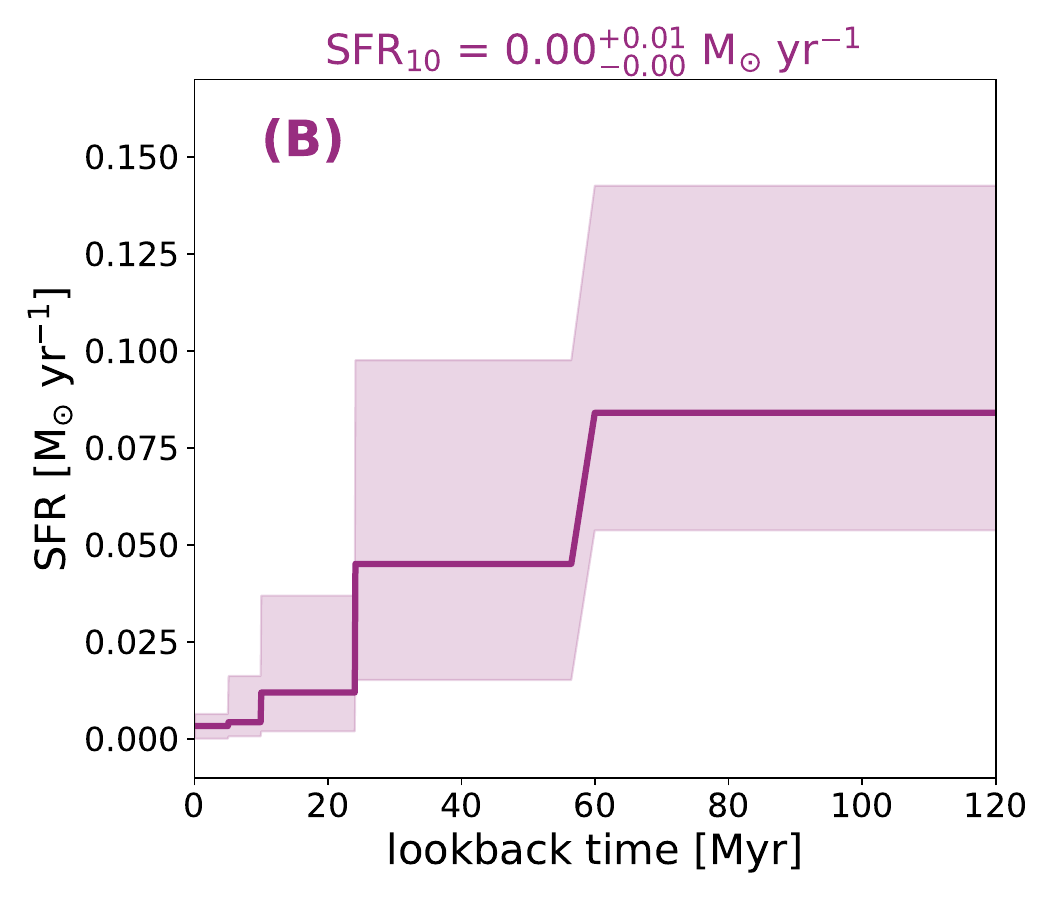}
    \includegraphics[width=0.3\linewidth]{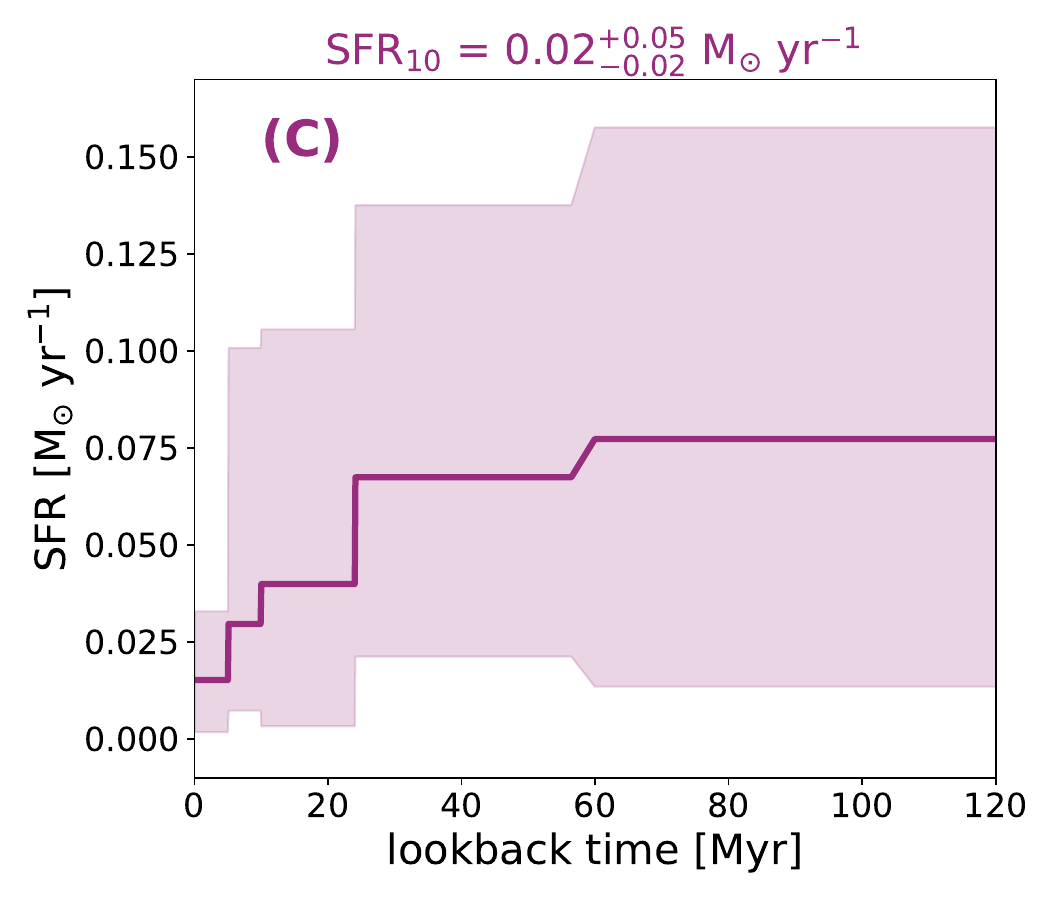}
    \caption{Retrieved star formation histories for a galaxy with no recent star formation, at $z=3$, and stellar mass of log(M$_{\star}$/M$_{\odot}$)$=7.5$. The output SFR$_{10}$ is given in the title of each panel. From left to right, the panels show cases with decreasing amount of photometric information: (A) all JWST and HST bands, (B) JWST medium and wide bands, and (C) JWST wide bands only. In all cases, \texttt{Prospector} is able to reliably retrieve the shape of declining input SFH, albeit with increasing error bars, which reflect the amount of input information.}
    \label{app:SFH_z3}
\end{figure*}

\begin{figure*}
    \centering
    \includegraphics[width=0.3\linewidth]{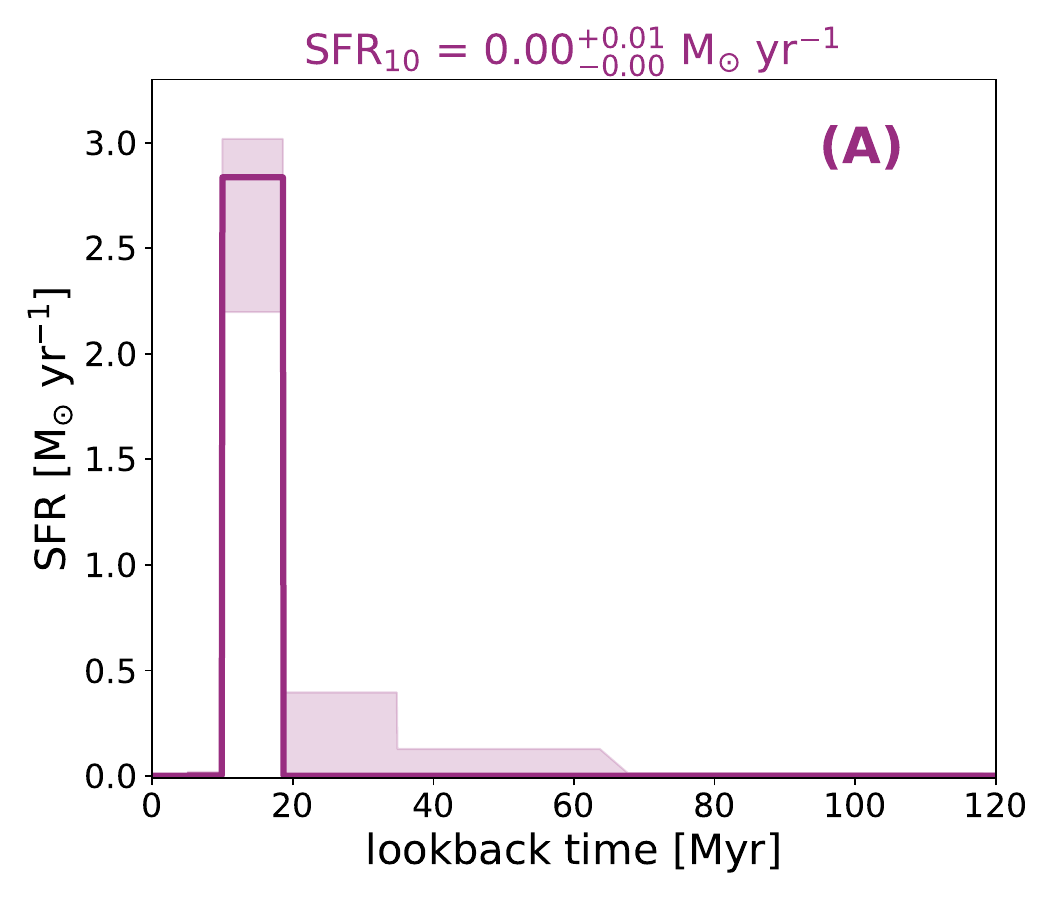}
    \includegraphics[width=0.3\linewidth]{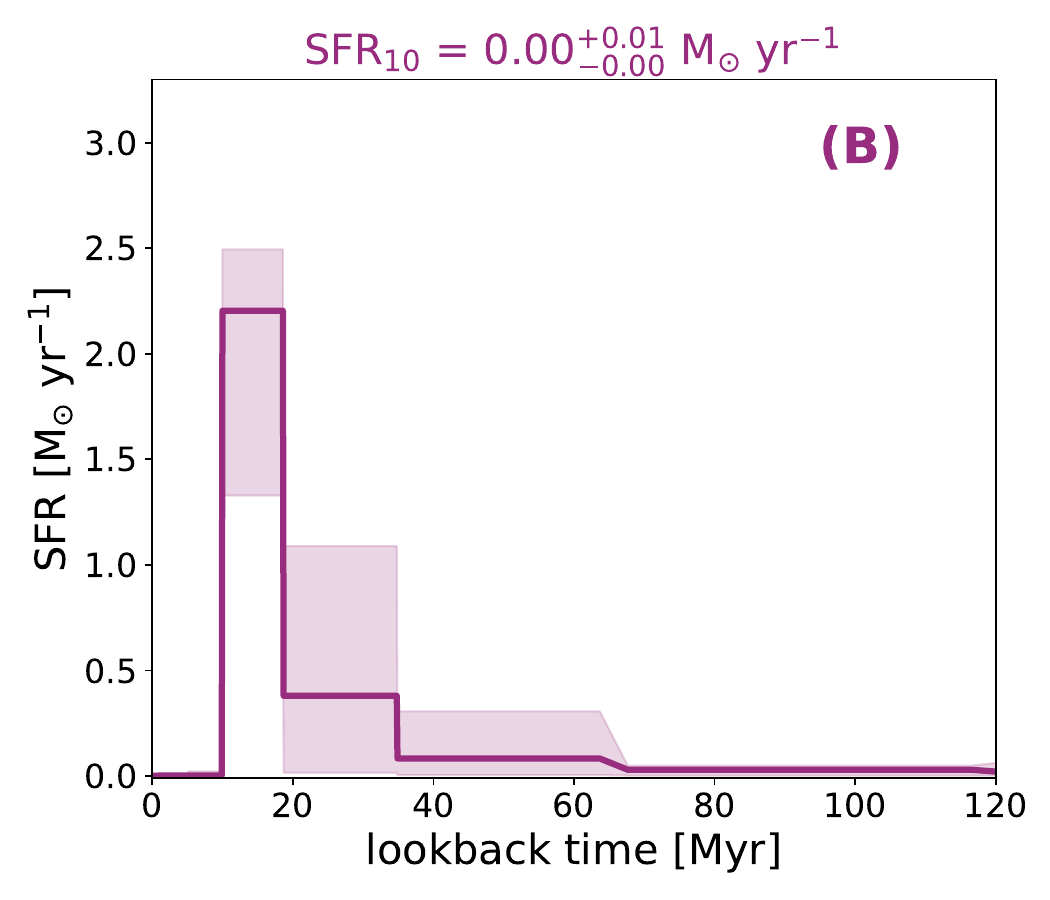}
    \includegraphics[width=0.3\linewidth]{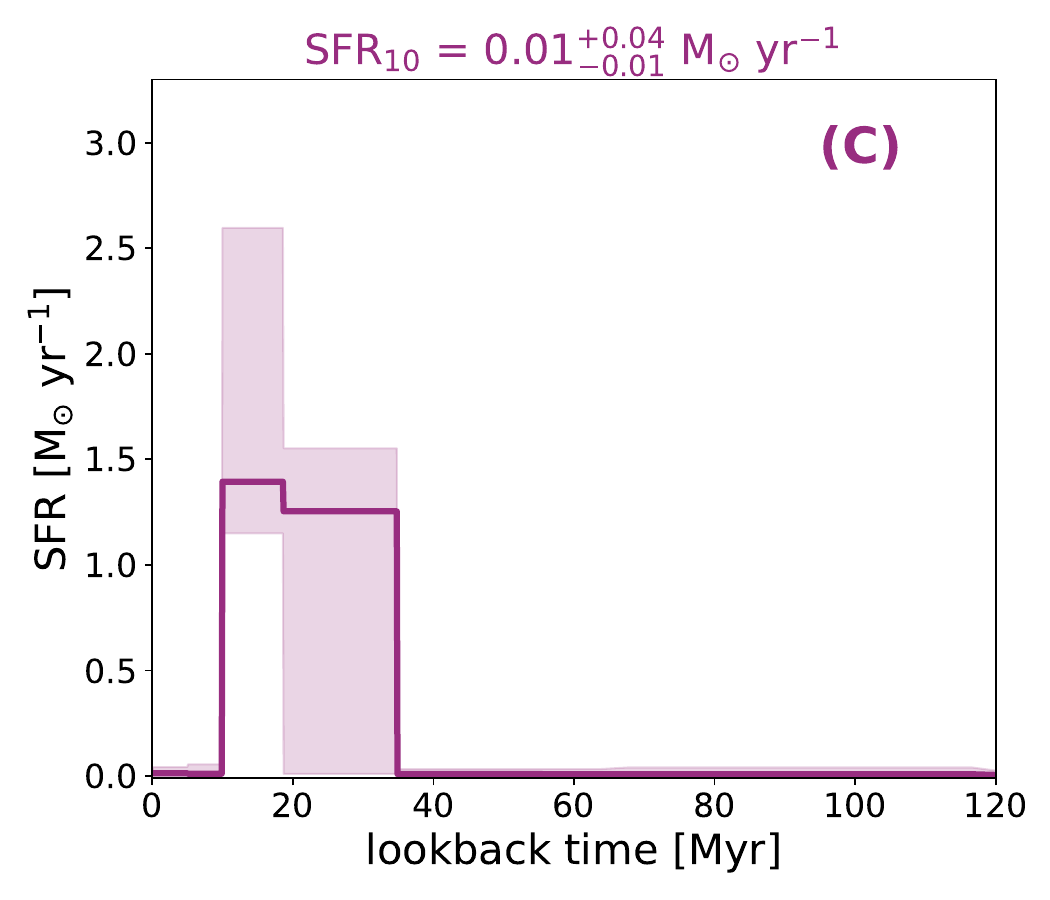}
    \caption{Same as Figure~\ref{app:SFH_z3} but for a galaxy at $z=8.35$.}
    \label{app:SFH_z8}
\end{figure*}
To assess the limitations of \texttt{Prospector} when retrieving SFHs, and consequently, SFRs, we use as input the best-fit photometry of two galaxies in our sample with no recent star formation, low stellar mass (log(M$_{\star}$/M$_{\odot}$)$\approx$7.5) and at the extremes of our redshift range ($z=3.00$ and $z=8.35$). The input photometry and SFHs used are shown in Figure~\ref{app:SFH_input}. The objective of this test is to confirm that \texttt{Prospector} can retrieve SFHs and SFRs reliably at low stellar masses (below the stellar mass completeness limit derived in Section~\ref{sec:completeness}).

We define three cases for each galaxy, depending on the number of photometric bands used in the fitting routine: 
(A) all HST and JWST bands potentially available in our data, (B) all JWST bands that are potentially available, including both medium and wide bands, and (C) only JWST wide bands.

In Figures~\ref{app:SFH_z3} and ~\ref{app:SFH_z8} we show the output SFHs for the cases described above, for $z=3.00$ and $z=8.35$, respectively. It can be seen that at both redshifts, the output SFHs reproduce the shape of the input SFHs. In turn, this means that the stellar masses and SFR$_{10}$ are also well retrieved. Importantly, this behaviour remains true as we reduce the number of photometric bands used as input. Although, as expected, the error bars increase when less information is available. 
With this simple test, we conclude that \texttt{Prospector} can indeed retrieve low SFRs at all the redshifts studied in this work, for galaxies with low stellar masses. Moreover, it can retrieve them reasonable well even in cases where no medium band observations are available.

\section*{Affiliations}
\noindent
{\it
$^{1}$The Kavli Institute for Cosmology (KICC), University of Cambridge, Madingley Road, Cambridge, CB3 0HA, UK\\
$^{2}$Cavendish Laboratory, University of Cambridge, 19 JJ Thomson Avenue, Cambridge, CB3 0HE, UK\\
$^{3}$Centre for Astrophysics Research, Department of Physics, Astronomy and Mathematics, University of
Hertfordshire, Hatfield AL10 9AB, UK\\
$^{4}$Steward Observatory, University of Arizona, 933 N. Cherry Avenue, Tucson, AZ 85721, USA\\
$^{5}$Center for Astrophysics $|$ Harvard \& Smithsonian, 60 Garden St., Cambridge MA 02138 USA\\
$^{6}$Department of Astronomy and Astrophysics, The University of Chicago, Chicago, IL 60637 USA\\
$^{7}$Kavli Institute for Cosmological Physics, The University of Chicago, Chicago, IL 60637 USA\\
$^{8}$Enrico Fermi Institute, The University of Chicago, Chicago, IL 60637 USA\\
$^{9}$Department of Astronomy and Astrophysics University of California, Santa Cruz, 1156 High Street, Santa
Cruz CA 96054, USA\\
$^{10}$NRC Herzberg, 5071 West Saanich Rd, Victoria, BC V9E 2E7, Canada\\
$^{11}$DARK, Niels Bohr Institute, University of Copenhagen, Jagtvej 155A, DK-2200 Copenhagen, Denmark\\
$^{12}$Institute of Astronomy, University of Cambridge, Madingley Road, Cambridge CB3 0HA, UK\\
$^{13}$European Space Agency (ESA), European Space Astronomy Centre (ESAC), Camino Bajo del Castillo
s/n, 28692 Villanueva de la Cañada, Madrid, Spain\\
$^{14}$Department of Physics, University of Oxford, Denys Wilkinson Building, Keble Road, Oxford OX1 3RH, UK\\
$^{15}$Scuola Normale Superiore, Piazza dei Cavalieri 7, I-56126 Pisa, Italy\\
$^{16}$European Southern Observatory, Karl-Schwarzschild-Strasse 2, 85748 Garching, Germany\\
$^{17}$Department of Physics and Astronomy, University College London, Gower Street, London WC1E 6BT, UK\\
$^{18}$Department of Astronomy, University of Wisconsin-Madison, 475 N. Charter St., Madison, WI 53706, USA\\
$^{19}$Centro de Astrobiología (CAB), CSIC-INTA, Ctra. de Ajalvir km 4, Torrejón de Ardoz, E-28850, Madrid,
Spain\\
$^{20}$NSF National Optical-Infrared Astronomy Research Laboratory, 950 North\\
}

\bsp	
\label{lastpage}
\end{document}